\documentclass[12pt]{article}

\usepackage[T1]{fontenc}
\usepackage[margin=1in]{geometry}

\usepackage{graphicx} 
\usepackage{authblk} 
\usepackage{url,hyperref,lineno,microtype,subcaption}
\usepackage{amsmath}
\usepackage{xcolor}
\usepackage{natbib}
\usepackage{wrapfig}
\usepackage{underscore}
\usepackage{gensymb}
\usepackage{amssymb} 
\usepackage{aas_macros} 
\usepackage{tikz,lipsum,lmodern}
\usepackage[most]{tcolorbox}

\begin{document}



\clearpage 
\thispagestyle{empty} 
\begin{tikzpicture}[remember picture, overlay]
    \node at (current page.center) {
        \includegraphics[width=\paperwidth, height=\paperheight]{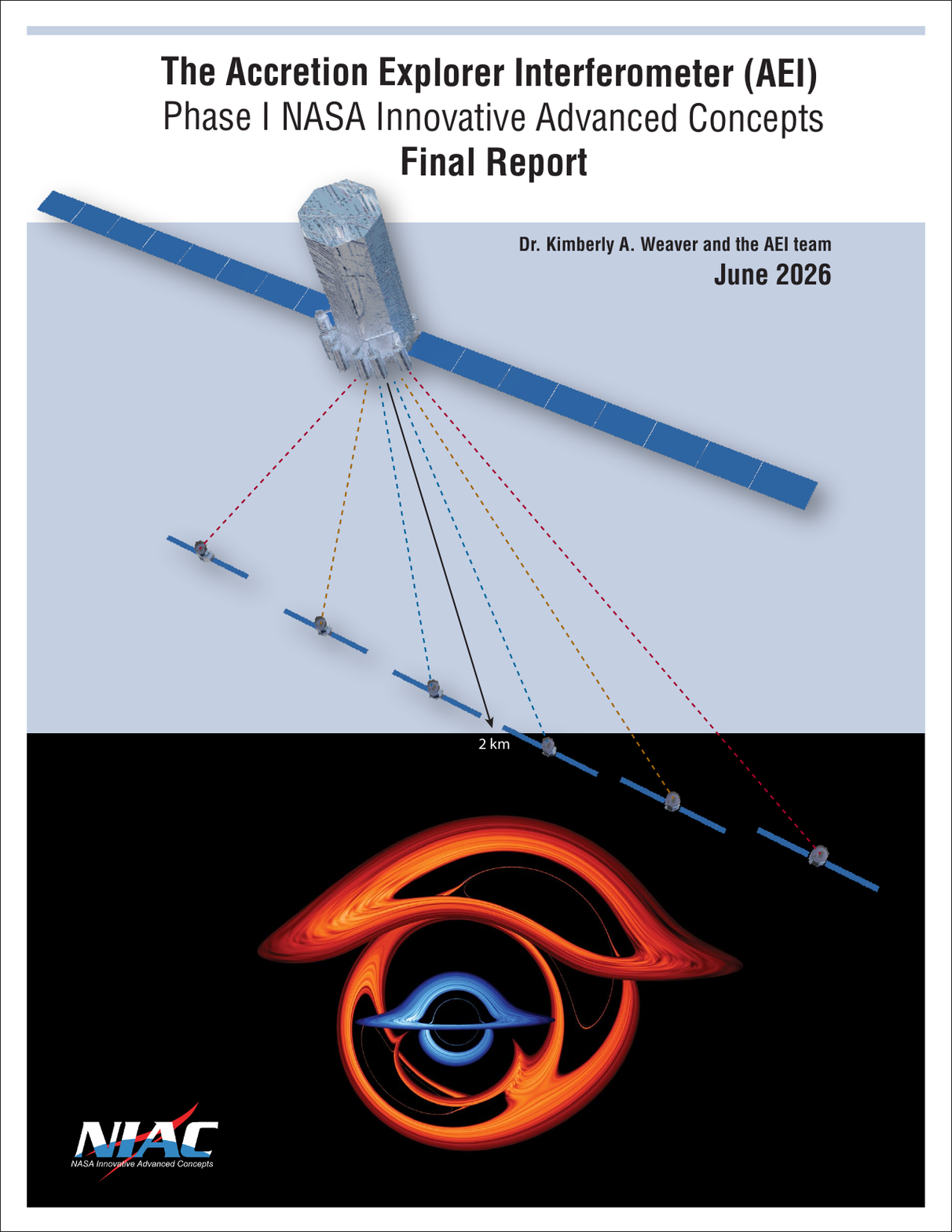}
    };
\end{tikzpicture}
\clearpage 


\author{First author \and second author \and third author}


\vspace{1cm}
{\bf The Accretion Explorer Interferometer NIAC Phase I Study Team}

\vspace{1cm}
{\bf Principal Investigator:} Kimberly A. Weaver (NASA/GSFC, Code 662)

\vspace{0.7cm}
{\bf Co-Investigators:} 

Jenna M. Cann (NASA/GSFC, Code 662; University of Maryland, Baltimore Co.)

Kenneth Carpenter (NASA/GSFC, Code 667)

Maurice Leutenegger (NASA/GSFC, Code 662)

Takashi Okajima (NASA/GSFC, Code 662)

Scott Rohrbach (NASA/GSFC, Code 551)

\vspace{0.5cm}
{\bf Mission Design Lab Engineers:}

Liz Matson (Integrated Design Center Manager; NASA/GSFC, Code 501)

Craig Stevens (Mission Design Lab Team Lead; NASA/GSFC, Code 533)

George Bussey (RF Communications; NASA/GSFC, Code 524)

Jean-Etienne Dongmo (ACS; NASA/GSFC, Code 534)

David Kim (Electrical; NASA/GSFC, Code 524)

Jess Lewis (Mechanical Systems; NASA/GSFC, Code 425)

Khashayar Parsay (Flight Dynamics; NASA/GSFC, Code 532)

Sharon Peabody (Thermal; NASA/GSFC, Code 523)

Alison Rao (Mission Systems, Propulsion; NASA/GSFC, Code 533)

Sara Riall (Mechanical Design; NASA/GSFC, Code 525)

Marta Shelton (Optical Communications; NASA/GSFC, Code 524)

Kan Yang (Instrument Design Lab Team Lead; NASA/GSFC, Code 520)

\vspace{0.5cm}
{\bf Collaborators:} 

Isabella Carlton (University of Utah)

Webster Cash (University of Colorado, Boulder)

Kaylee DeGennaro (Brown University)

Emma Kleiner (The City University of New York)

John Krizmanic (NASA/GSFC, Code 661)

Erini Lambrides (University of Maryland, College Park; NASA/GSFC)

Miranda McCarthy (The City University of New York)

Jeffrey McKaig (NASA Postdoctoral Fellow (ORAU); NASA/GSFC)

Atul Mohan (Catholic University; NASA/GSFC; East Asian ALMA Regional Center, NAOJ, Tokyo)

Teresa Monsue (Catholic University; NASA/GSFC Code 667)

Ryan Pfeifle (United States Naval Observatory)

Mainak Singha (Catholic University; NASA/GSFC)

Melville Ulmer (Northwestern University)

Laura D. Vega (University of Maryland, College Park; NASA/GSFC)

Kelly E. Whalen (NASA Postdoctoral Fellow (ORAU); NASA/GSFC)

\newpage

\tableofcontents

\begin{figure}
    \centering
    \includegraphics[width=0.7\textwidth]{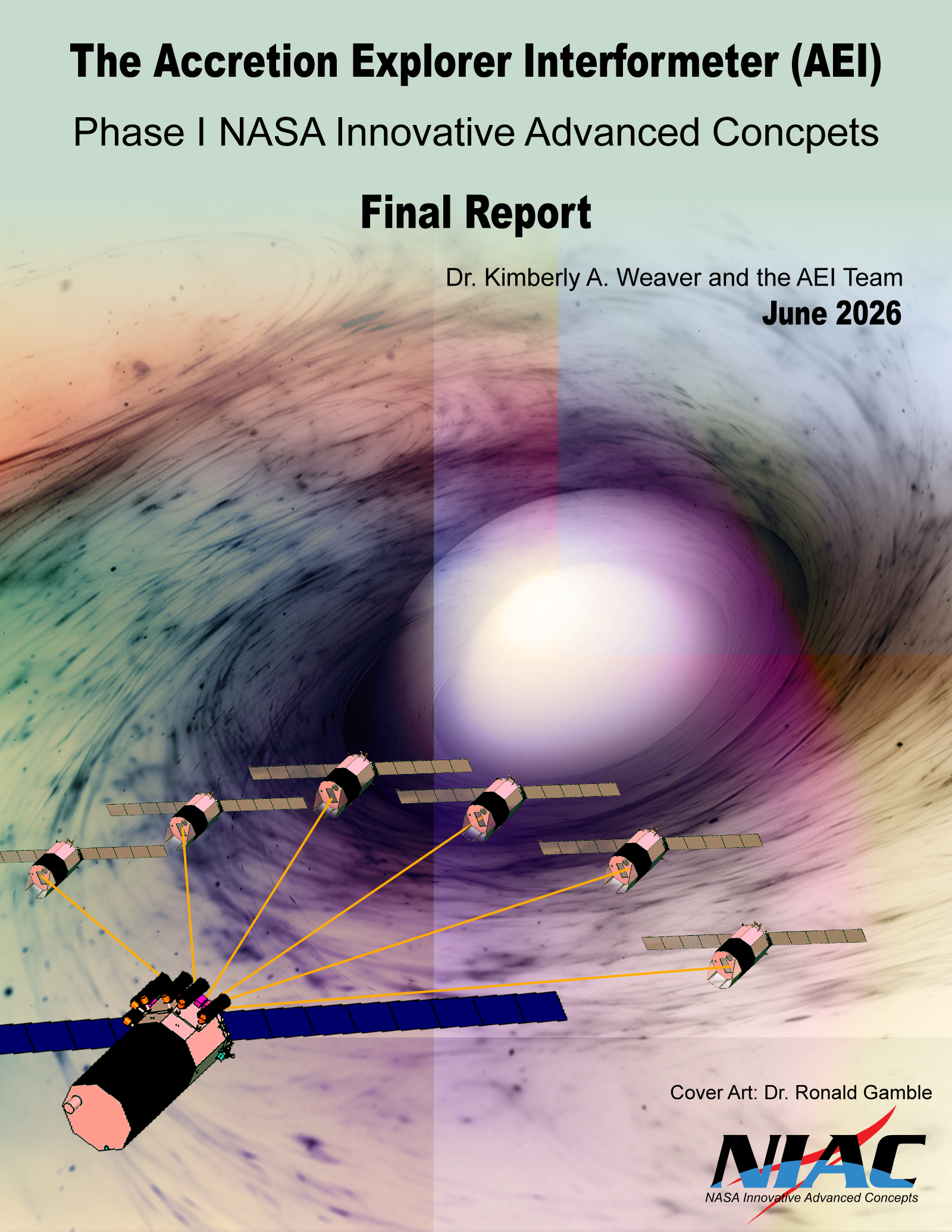}
\end{figure}

\newpage

\section{Executive Summary}

We must create superb X-ray images to understand the detailed physical processes behind some of the most powerful astronomical objects. The need to achieve this capability has been known for decades. But as time proceeds, X-ray astronomy falls further behind other wavebands that are steadily increasing their imaging capacity. Radio astronomy in particular has reached an angular resolution on the order of micro arcseconds ($\mu$as) via aperture synthesis interferometry, using the interference of electromagnetic waves from many small telescopes together to simulate having a much larger telescope. Developing an equivalent high-resolution capability in the X-ray band would be a game changer for high-energy astrophysics. We will understand how supermassive black holes (SMBH) grow and evolve. We will learn what powers astrophysical jets. We will learn how young, active stars affect the habitability of their planets. Technologically, our NIAC study has shown that the Accretion Explorer Interferometer (AEI) concept, unlike the original MAXIM concept, is more feasible in operation, being only 2 km long, versus $\sim450$ km. 
Our study has also shown that satellite station keeping is possible, leveraging from LISA pathfinder technology, and using large mirror flats plus an X-ray beamsplitter for enabling technology is feasible. Thus, a Phase II NIAC proposal is justified. Our Phase I proposal stated that {\it ...a NIAC Phase I study alone will challenge and change the conversation around future possibilities for NASA’s flagships and probe class missions.} We believe we have done this and have progressed a mission concept to a feasible architecture for further study to expand the possibilities around today's future X-ray missions.

\section{Introduction and Innovation}

\begin{wrapfigure}{r}{0.46\textwidth}
    \centering   \includegraphics[width=0.45\textwidth]{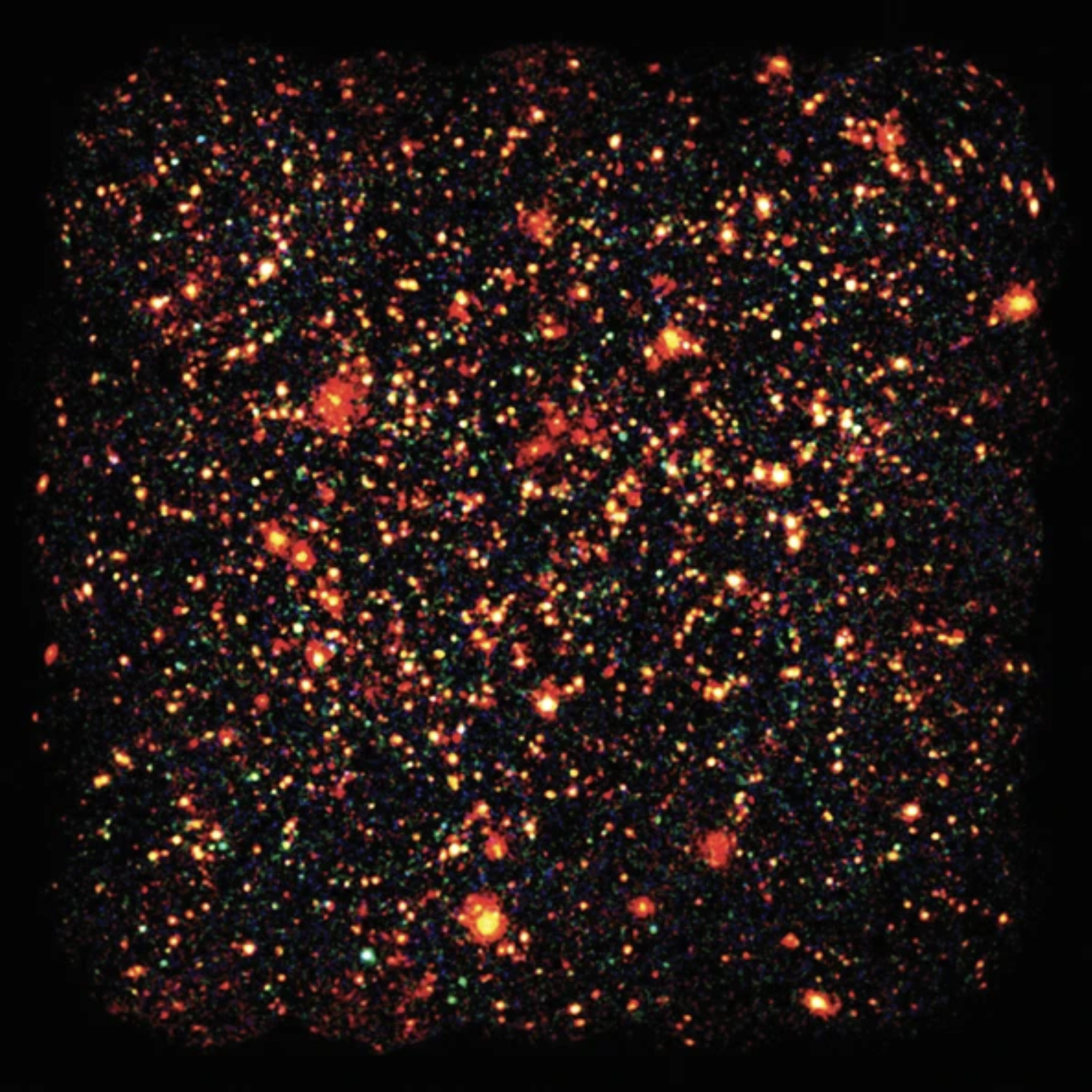}
    \caption{\small {\bf A Night Sky of black holes!} An XMM-Newton deep, wide ($>1$ deg$^2$) extragalactic X-ray survey \citep{2007ApJS..172...29H}.
    Most of the $\sim2,000$ point-like sources represent accreting supermassive black holes.
    }
      \label{fig:deepfield}
\end{wrapfigure}

We have studied an architecture for the Accretion Explorer Interferometer (AEI) concept, a mission that will achieve a major increase in observational capabilities by creating X-ray images up to six orders of magnitude better than the current premier X-ray imaging telescope, Chandra. AEI will resolve black hole accretion, enable discoveries related to the mysteries of black hole power (Figure \ref{fig:deepfield}), and capture images of powerful stellar flares.

AEI enables fundamentally new science with ultra-high resolution X-ray images and will revolutionize studies of black holes by providing clear X-ray images to match breakthrough $\mu$as radio images with the Event Horizon Telescope, but also probing a wider range of targets and cosmological distances. AEI addresses NASA priorities within the Astrophysics Vision/SMD Priorities Strategy 1.1/NASA’s Strategic Plan SO 1.2, which are to Understand the Sun, Solar System, and Universe: “How does the Universe work?”. Gaining a more complete understanding of SMBH and active galactic nuclei (AGN) also supports the Decadal Survey on Astronomy and Astrophysics 2020 (hereafter Astro2020, \cite{national2021decadal}) science themes of “New Messengers and New Physics” and “Cosmic Ecosystems.” Astro2020 explicitly calls out that our “rudimentary” knowledge of “how [AGN] spectra, jets, and winds vary with luminosity, black hole mass, black hole spin, etc. [is] a major bottleneck” in understanding galaxy evolution. We also address Astro2020’s Habitable Worlds objective that calls for “A high spatial and spectral resolution X-ray space observatory to probe stellar activity across the entire range of stellar types, including host stars of potentially life-sustaining exoplanets.”  

\begin{figure}[h!]
\centering
  \begin{subfigure}[b]{0.95\textwidth}
 \includegraphics[width=\textwidth]{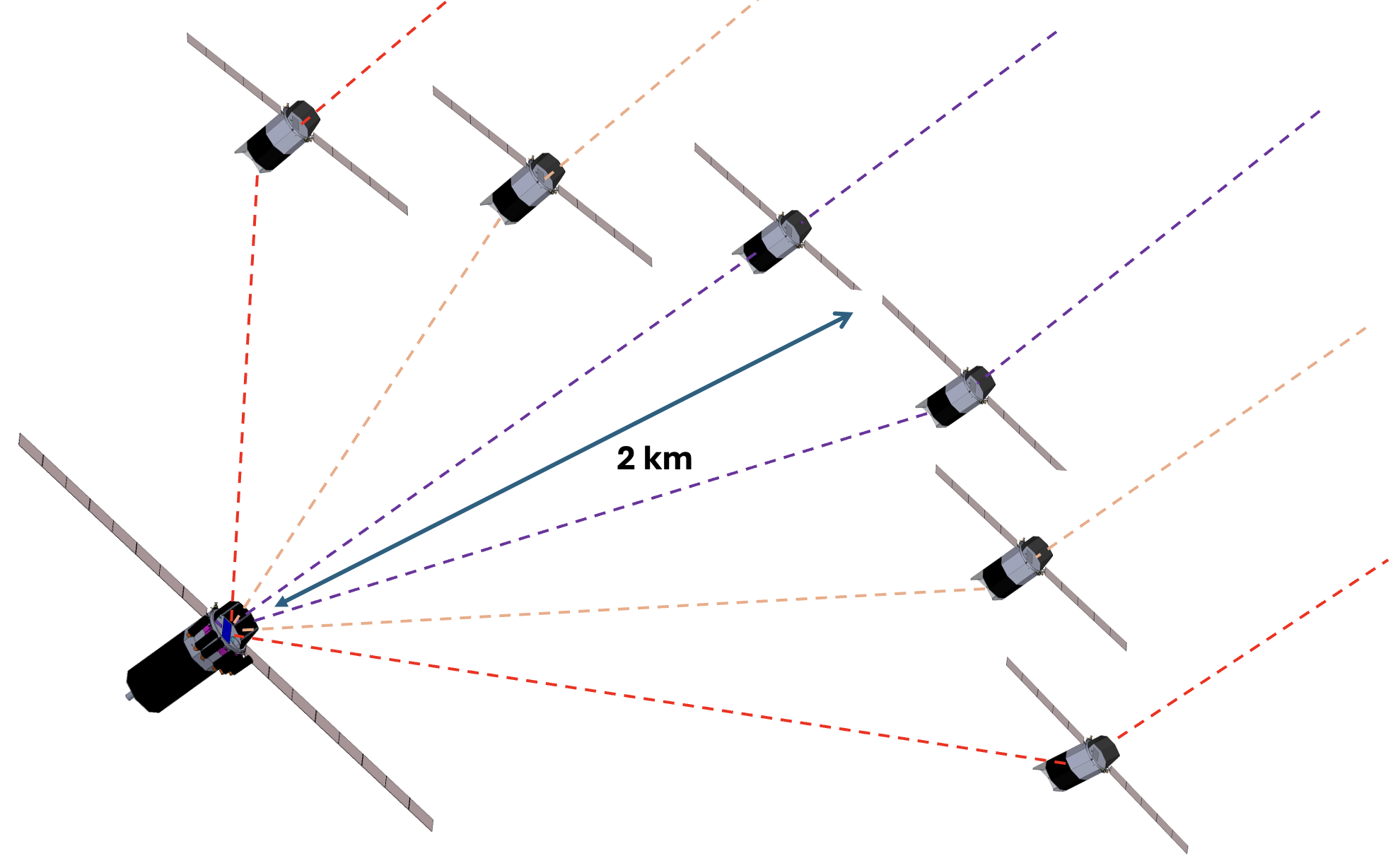}
     \caption{Deployed AEI array at L2. The satellite that holds the detectors and combiner optic is to the lower left and the other six smaller satellites contain mirror flats to collect and redirect X-rays from the target (indicated by dashed lines).}
  \end{subfigure}
  
\bigskip

  \begin{subfigure}[b]{0.95\textwidth}
\includegraphics[width=\textwidth]{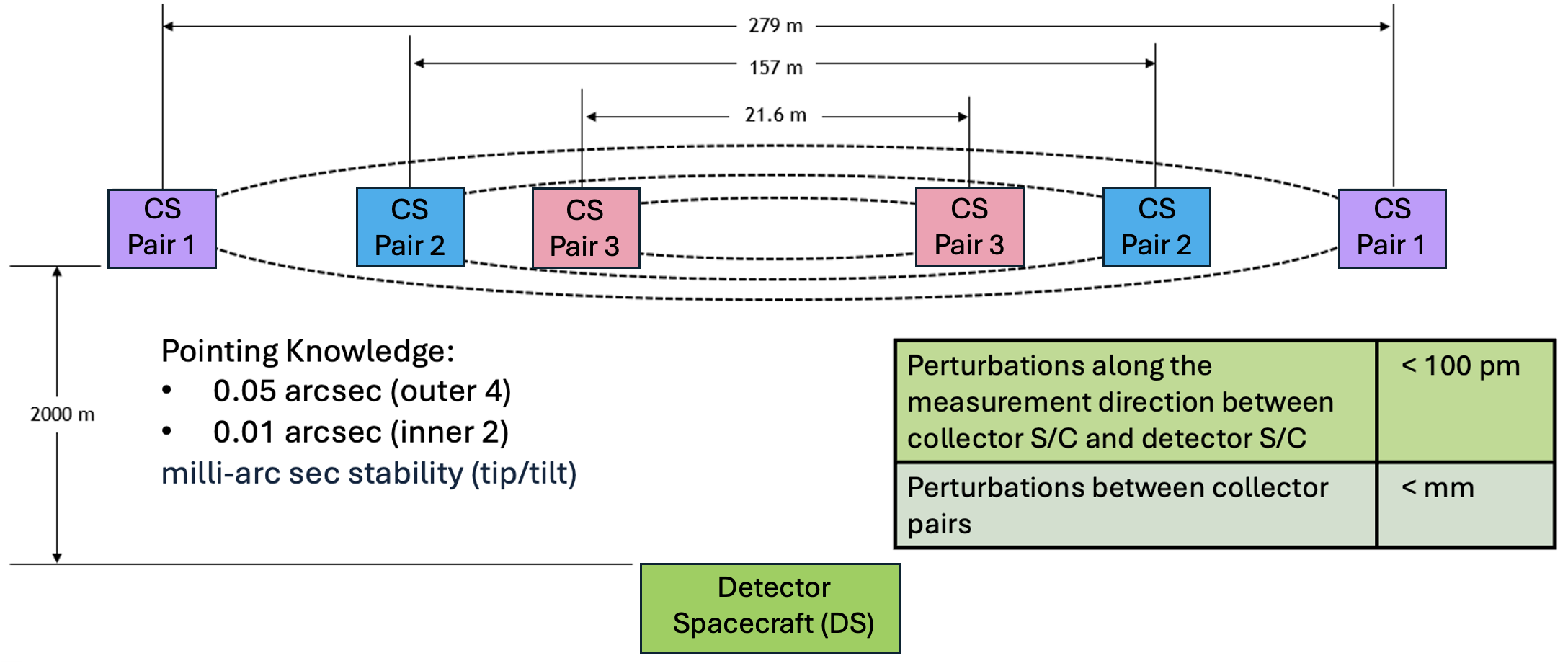} 
    \caption{Array alignment chart.}
  \end{subfigure}
    \caption{\small {\bf The AEI operational array configuration.} (a) The conceptual idea of the deployed array showing the light paths from collectors to the detector. (b): a visual chart showing all of the relative distances between the spacecraft, including baselines, and listing the needed stabilities for maintaining the required pathlength differences that would allow the configuration to maintain phase information and produce X-ray fringes.
}
   \label{fig:constellation} 
\end{figure}

This NIAC study is also forward-looking toward new science goals for the 2030 Astronomy and Astrophysics Decadal survey, as well as reaching backward to fulfill an identified need. We note that a Black Hole Imager was called out within the Visionary era of Technologically Advanced Missions described by NASA’s Enduring Quests, Daring Visions (2013, Astrophysics Roadmap) to spatially resolve SMBH and their environs and study the effects of extreme gravity. Specifically, an X-ray interferometer with (sub) $\mu$as resolution is discussed as a visionary notional mission, to complement an X-ray Surveyor. Here we present an X-ray interferometer concept that could form the core of such a flagship-sized mission for the future (Figure \ref{fig:constellation}).



  

An interferometer for X-rays has been studied before, and we did not seek to replicate that work. In particular, descriptions of mirror flat alignments and tolerances have been calculated from engineering designs by \cite{cash2003a, cash2005imaging, gendreau2004, 2003SPIE.4851..568S} - MAXIM,  and \cite{uttley2020, Uttley2021, denhartog2020}. Those tolerances will apply here, and since our baseline design for a large interferometer is simpler (6 large mirrors instead of 32 in the case of MAXIM), we assume the same will apply. Instead, we sought and found a practical architecture that can achieve the following innovations: (1) the needed pointing on target, (2) the needed stability for required long exposures, (3) a reasonable science mission lifetime, and (4) testing the boundaries for necessary development. Technology development is still required to bring a mission like this to life, but our study has produced a feasible concept and thus a technology path forward.

\section{Background for the NIAC AEI study}
\begin{wrapfigure}{r}{0.45\textwidth}
    \centering
\includegraphics[width=0.43\textwidth]{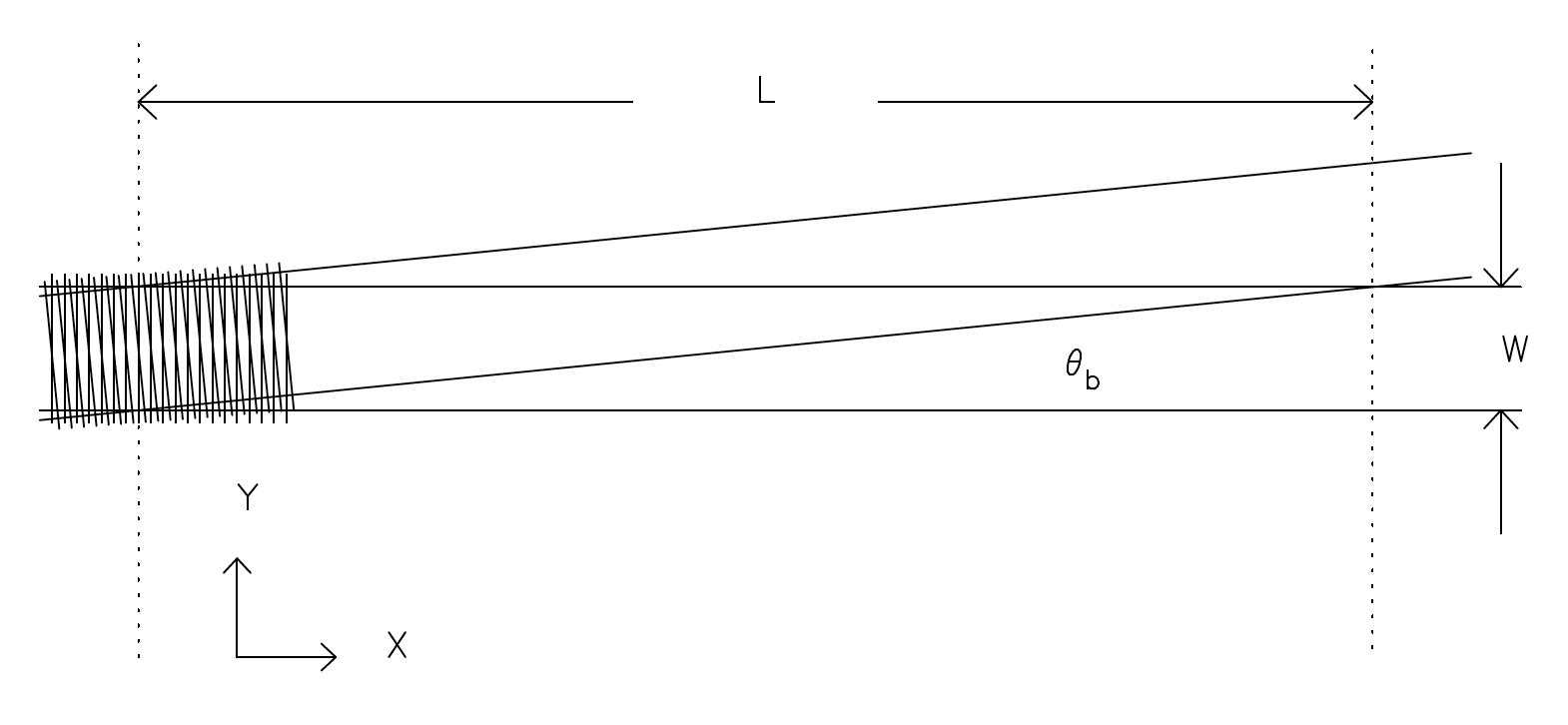}
\includegraphics[width=0.43\textwidth]{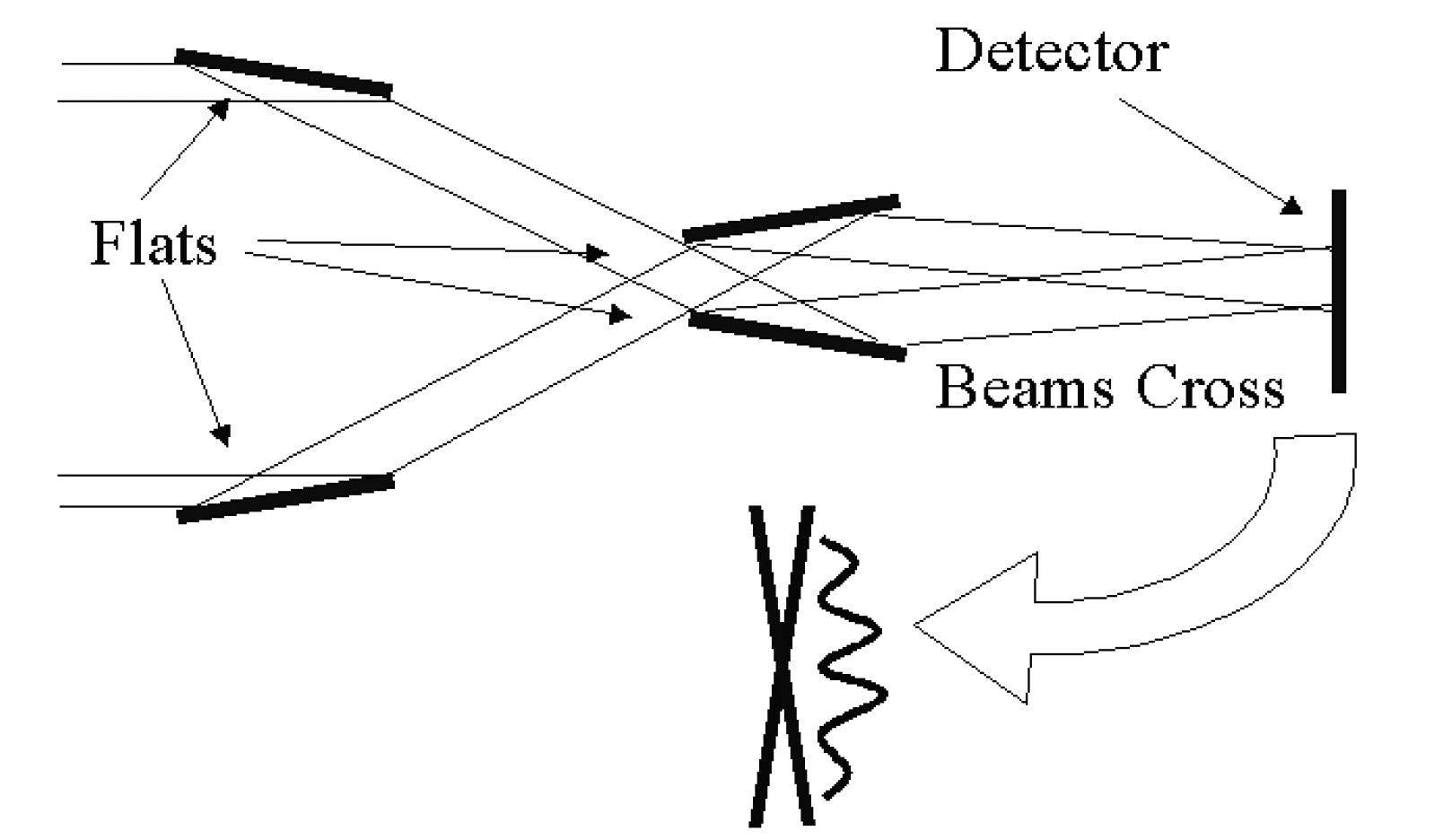}
    \caption{\small {\bf Parallel beams intersect to make interference fringes.} Top: From \cite{willingale2004}. Bottom, the original "x" configuration of mirror flats for the MAXIM design from \cite{cash2003a}.}
      \label{fig:fringes}
\end{wrapfigure}

 X-ray interference fringes were first demonstrated in the lab by \cite{cash2000}. Since then, designs have been put forward for space-based telescopes that depend on the concept of X-ray interferometry to reach ultra-high angular resolution. The basic idea of interferometry, which relies on the geometry of parallel beams intersecting to make interference fringes, is shown in Figure \ref{fig:fringes}. If the angle between the beams is $\theta_b$, and for a wavelength $\lambda$, then the fringe spacing is given by $\Delta$$y=\lambda/\theta_b$. In this example, $\theta_b=W/L$ and the number of fringes, $N_f$, is given by $N_f=W/\Delta$$y$ \citep{willingale2004}. Four flat X-ray mirrors can be used at grazing incidence angles of $\theta_g\sim2^{\circ}$ to focus incoming beams of X-rays to a common plane. Depending on the relative placement of these mirrors, the physical length of the telescope changes. The field of view of the system depends on the combination of the sizes of the reflecting mirrors and the detector area.

\newpage

A NIAC study for the Micro-arcsecond X-ray Imaging Mission (MAXIM) black hole imager was led by Webster Cash in the early 2000’s \citep{cash2005imaging}, and tests were performed in the lab \citep{gendreau2004}. \cite{cash2003a} and \cite{cash2005imaging} outlined and promoted an architecture for a diffraction limited structure with a 100 m baseline and 450 km focal length. NASA did not adopt those plans, and the technology unfortunately had time to languish. More recently, \cite{uttley2020} proposed an X-ray imager to the European Space Agency (ESA: Voyage 2050) with a compact optical design \citep{willingale2004}. A testbed was considered \citep{denhartog2020}, however this mission was not adopted at the time. We note that X-ray interferometry is still being considered as a future possibility by ESA.

We believe MAXIM was a groundbreaking study, and now is the time to revisit and extend it after many years of technology evolution, including advancements made in formation flying \citep{monnier2019}, and the pointing precision and stability demonstrated with the James Webb Space Telescope (JWST) and the Laser Interferometer Space Antenna (LISA) pathfinder. Our architecture is substantially different from MAXIM, envisioning (eventually) larger effective areas per energy channel with a combination of large collecting mirror area, high combiner optic throughput and large-area CCD detectors. Our combiner optic eliminates the long focus lengths of the MAXIM and original Uttley designs, dramatically decreasing the distances between the formation-flying mirror to detector spacecraft that were previously required and the distance between the combiner optic and the detector focal plane, thus allowing us to reach $~\sim\mu$as imaging resolution with a comparably smaller system.

In the Cash design, the interferometer consists of two rings of flat mirrors.\footnote{\href{https://www.niac.usra.edu/files/studies/final_report/389Cash.pdf}{https://www.niac.usra.edu/files/studies/final_report/389Cash.pdf}}
Each ring contains 32 flat mirrors, and they published a mirror error budget tree for their ``periscope'' X-ray telescope, which was an evolution of their "x" configuration (Figure \ref{fig:fringes}), as well as mirror tolerances and tradeoffs \citep{2003SPIE.4851..568S}.\ They point out a crucial concern with jitter on the focal plane, and their recommendation at the time was to explore propulsion systems with microthrusters (providing force in the small micronewton ($\mu$$N$) to millinewton ($mN)$ range). Other areas needing attention included (1) interactions between the array elements, (2) thermal stability for the spacecraft, and (3) specific microthruster performance. We note that the first set of mirror flats in Figure \ref{fig:fringes} (bottom) can be connected to the job of collector satellites in our design, while the second set of flats and the detector plane essentially become the detector satellite in our design but with a different alignment (Section \ref{combiner}).

\cite{uttley2020, Uttley2021} explore a more compact interferometer that can fit onto a single spacecraft and would reach $\sim100$ $\mu$as resolution. In addition to discussing pointing stability and telescope jitter, they also mention the importance of achieving thermal stability for the spacecraft. Finally they calculate exposure times needed to reach astrophysical flux levels. We have taken the approach in our NIAC Phase I study to examine the majority of the outstanding issues from both the Cash and Uttley investigations to find practical solutions. We address formation flying, pointing precision, a partial thermal assessment, and the required exposure times and how all of these things feed into the propulsion budget, and thus the lifetime, of an operational system. We also describe the individual resolutions required for X-ray spotting telescopes needed for target acquisition and centering, and differentiate them from the full-array $\mu$as interferometer. 

\begin{tcolorbox}[colback=green!5!white,colframe=green!60!black,title=In This NIAC Report for a Stable Formation Array We Explore:]
  \begin{itemize}
       \item Microthruster use and propulsion budget (Section \ref{mdl} and \ref{thrusters})
   \item Satellite modular approach and future adaptability (Section \ref{modular})
      \item Formation flying maintenance and array stability for $\mu$as resolution (Section \ref{attitudecontrol})
      \item Target acquisition and observations planning requirements (Section \ref{targets})
      \item  Communications and data transfer considerations (Section \ref{comms})
  \end{itemize}
\end{tcolorbox}

\section{Goals of the Phase I study}
\label{goals}

We addressed well known technical challenges from prior X-ray interferometry designs of formation flying, pointing accuracy, stability, and observations feasibility with innovations that will make this a practical architecture for a future large mission such as a flagship. Our goals were as follows:

{\bf Goal 1:} Design a set of high level science goals and measurements to craft notional mission descriptions that feed into a Science Traceability Matrix (STM) for X-ray interferometry -- see Section \ref{stm}. We used our notional mission descriptions to design a strawperson observation schedule with viable targets that tested the limits of observation planning and the propulsion required for this architecture, to determine where technology improvements are required.


{\bf Goal 2:} Design a multiple-spacecraft system that aligns individual mirror pair baseline groupings provided by individual “collector mirror” spacecraft, with the pointing precision to achieve  $\mu$as resolution 
This was achieved with a Mission Design Lab study (Section \ref{mdl}).

{\bf Goal 3:} Assess whether this architecture can meet the required pointing stability on target. The visibility of fringes are significantly reduced if the variations in optical path difference of the combining wave trains become too large compared to the X-ray wavelength (e.g., \cite{uttley2020}), so a stability on the order of nm is needed across the length of an observation, immune to spacecraft vibrations. The LISA technologists have developed extremely precise measurements with a laser interferometer -- measuring relative distance between two test mirrors with an accuracy of a few picometers per square root Hertz \citep{armano2021}. We incorporate this technology into our design (Section \ref{lisa}). 

\begin{figure}
\centering
  \includegraphics[width=0.54\textwidth]{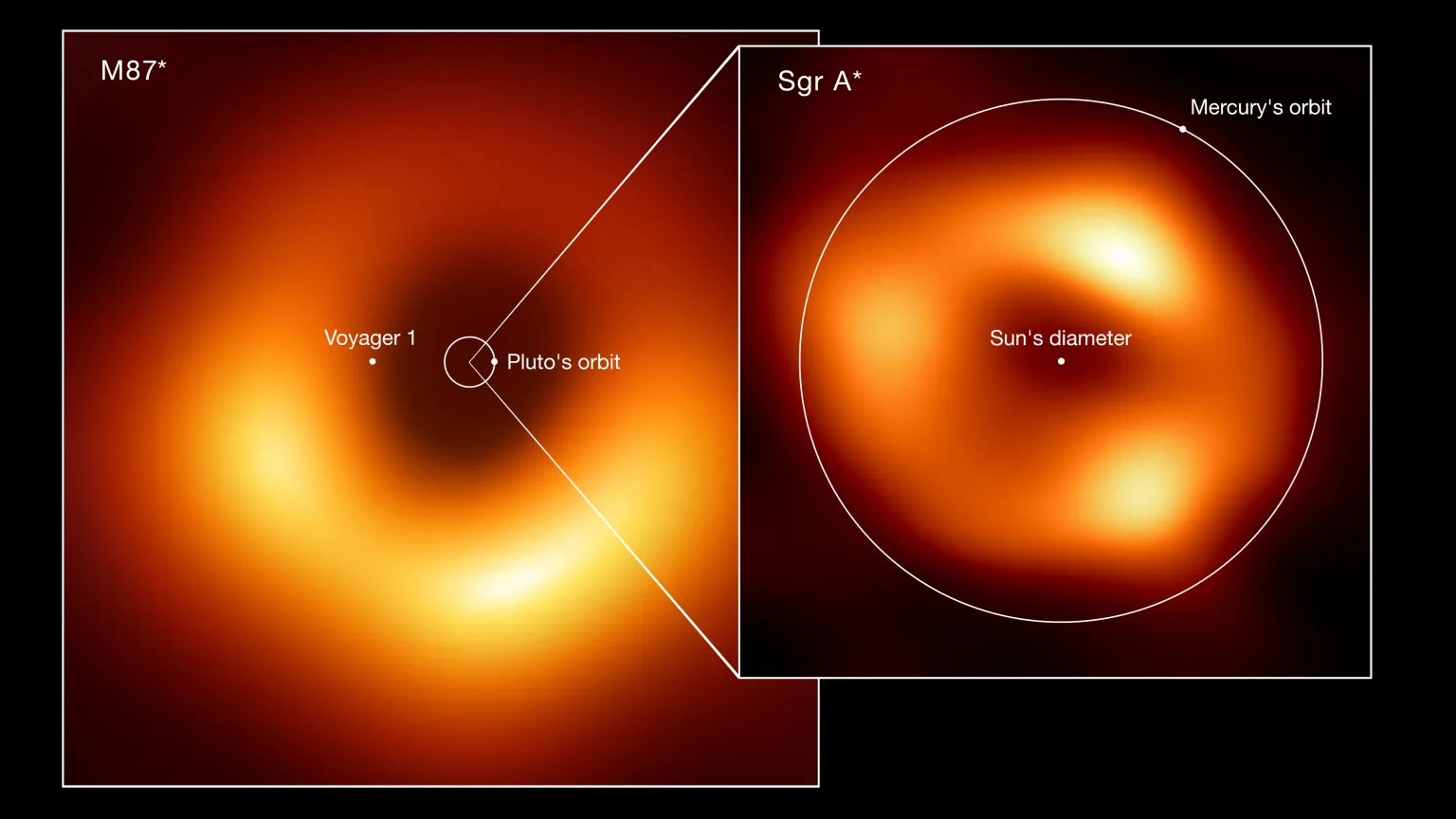}
  \includegraphics[width=0.45\textwidth]{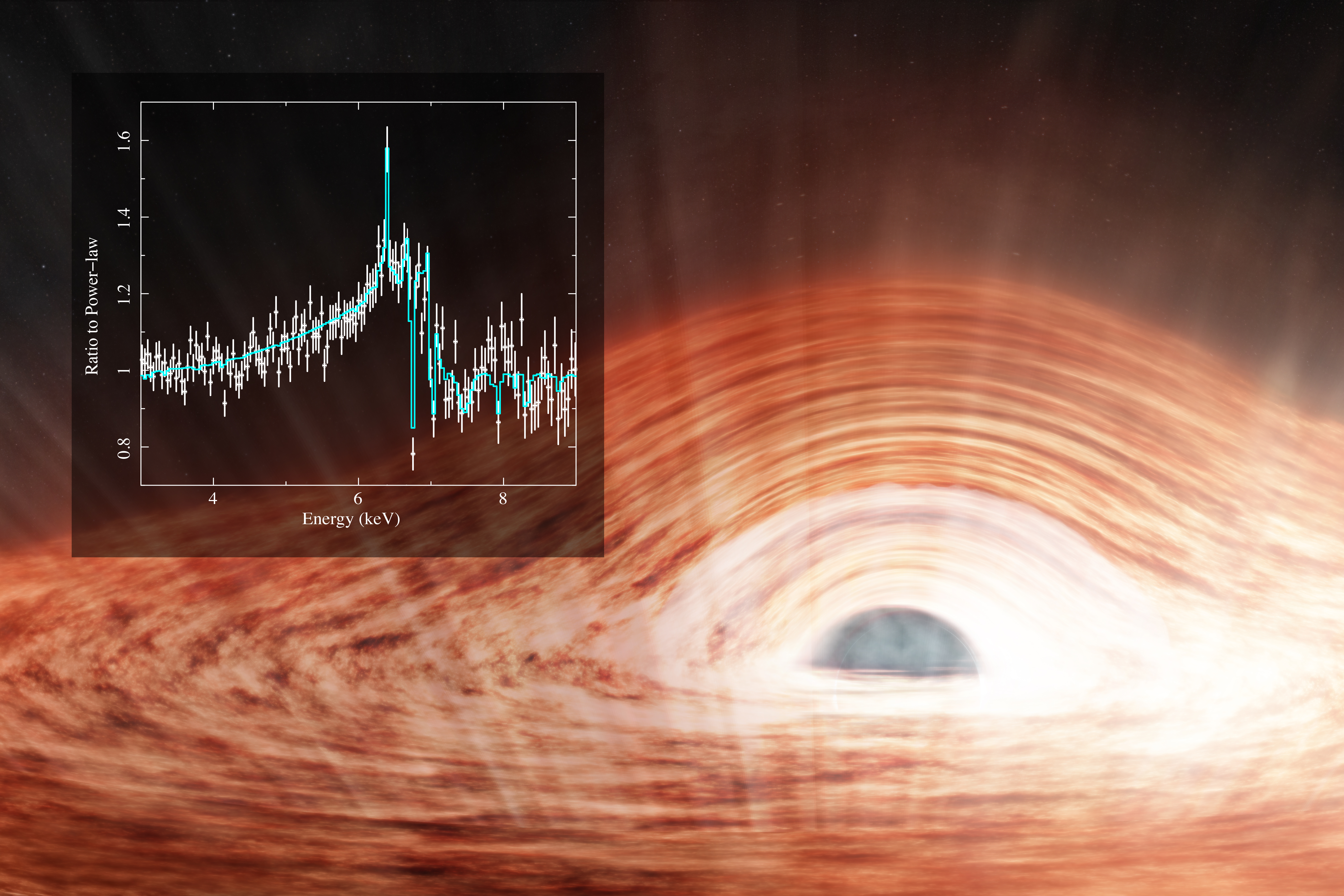}
    \caption{\small 
{\bf Compelling science stories about supermassive black holes made headlines across the globe.} Left: This picture shows a reconstructed image of the very innermost regions surrounding the nearby SMBH in M87 (left) and Sagittarius A* (right).\protect\footnotemark
 The images capture the glow of long wavelength radio light tracing the black hole “shadow.” But to confirm theories about black hole accretion, we need to capture much more powerful X-ray light from objects like M87*. Right: XRISM and NuSTAR together show the X ray spectrum along with an artist's rendition of material from a SMBH with relativistic effects that distort the appearance of the the strong Fe K$\alpha$ emission line (inset, \citep{2025ApJ...995..200B} and \protect\footnotemark). However, we need ultra-high resolution to truly image the distorted accretion disk in X-ray light.}
    \label{fig:M87} 
\end{figure}
\footnotetext{\href{https://www.eso.org/public/news/eso2406/}{https://www.eso.org/public/news/eso2406/}}
\footnotetext{\href{https://www.cfa.harvard.edu/news/new-x-ray-space-telescope-gives-sharpest-ever-glimpse-growth-rapidly-spinning-black-hole}{https://www.cfa.harvard.edu/news/new-x-ray-space-telescope-gives-sharpest-ever-glimpse-growth-rapidly-spinning-black-hole}}

{\bf Goal 4:} Determine the practicality of formation flying with small baselines. Small baselines between satellites in space are a risk (Monnier et al. 2019), and this affects specifically our inner pair of collecting mirror satellites with a $\sim22$ m baseline. We assessed how or whether to mitigate this risk. One option is to connect the inner baselines with a deployable truss (Hoyt, et al. 2013) or tether, while the outer spacecraft are left as free flyers.

{\bf Goal 5:} Capture engineering benefits that can contribute to other space missions. These include our study of very high levels of pointing precision for space-based interferometers, microthrusters, and propulsion budgets. We also captured engineering challenges and areas that require future development (Sections \ref{challenges}, \ref{technology}), to eventually feed into a technology roadmap.

\section{AEI Science}

AEI will enable fundamentally new science and will revolutionize studies of black holes by providing clear X-ray images to match breakthrough $\mu$as radio images with the Event Horizon Telescope (EHT - see Figure \ref{fig:M87}).
A preeminent NASA theme is that ``NASA explores the secrets of the Universe for the benefit of all.'' Some of the Universe’s most enigmatic secrets are supermassive black holes (SMBHs), verified in the 1990s with the Hubble Space Telescope (HST). Remarkably, even after decades, scientists don’t know how these light-bending objects produce their millions to billions of Suns worth of power. Such an understanding demands that we observe short wavelengths (X-rays: E $\sim0.5-20$ keV). However, our only window into this extreme region is via unresolved images, since the state-of-the-art resolution provided by the Chandra X-ray telescope is just $\sim~0.5-1''$, much courser than is needed to image regions near SMBHs where their most energetic behavior occurs. The black hole diameter is a miniscule $\sim60~\mu$as across even for the closest X-ray luminous SMBH, M87 \citep[Figure \ref{fig:M87} and][]{Lu2023}.

For a general discussion of high resolution X-ray imaging, see \cite{weaver2026}.
Our NIAC team examined three science pillars to assess the science case for an X-ray interferometer reaching $\mu$as resolution:

{\bf Science Case 1: Accretion.} The primary science objective of AEI is to observe the accretion phenomena around SMBH to directly measure their sources of extreme power. We require X-ray imaging up to 6 orders of magnitude better than current missions (i.e., Chandra) and even $\sim5$ orders of magnitude better than recently proposed missions (e.g., Lynx, AXIS). This capability would enable, for the first time, the opportunity in the X-rays to spatially map the accretion disk structures. 

\begin{figure}[t]
\centering
\includegraphics[width=0.99\textwidth]{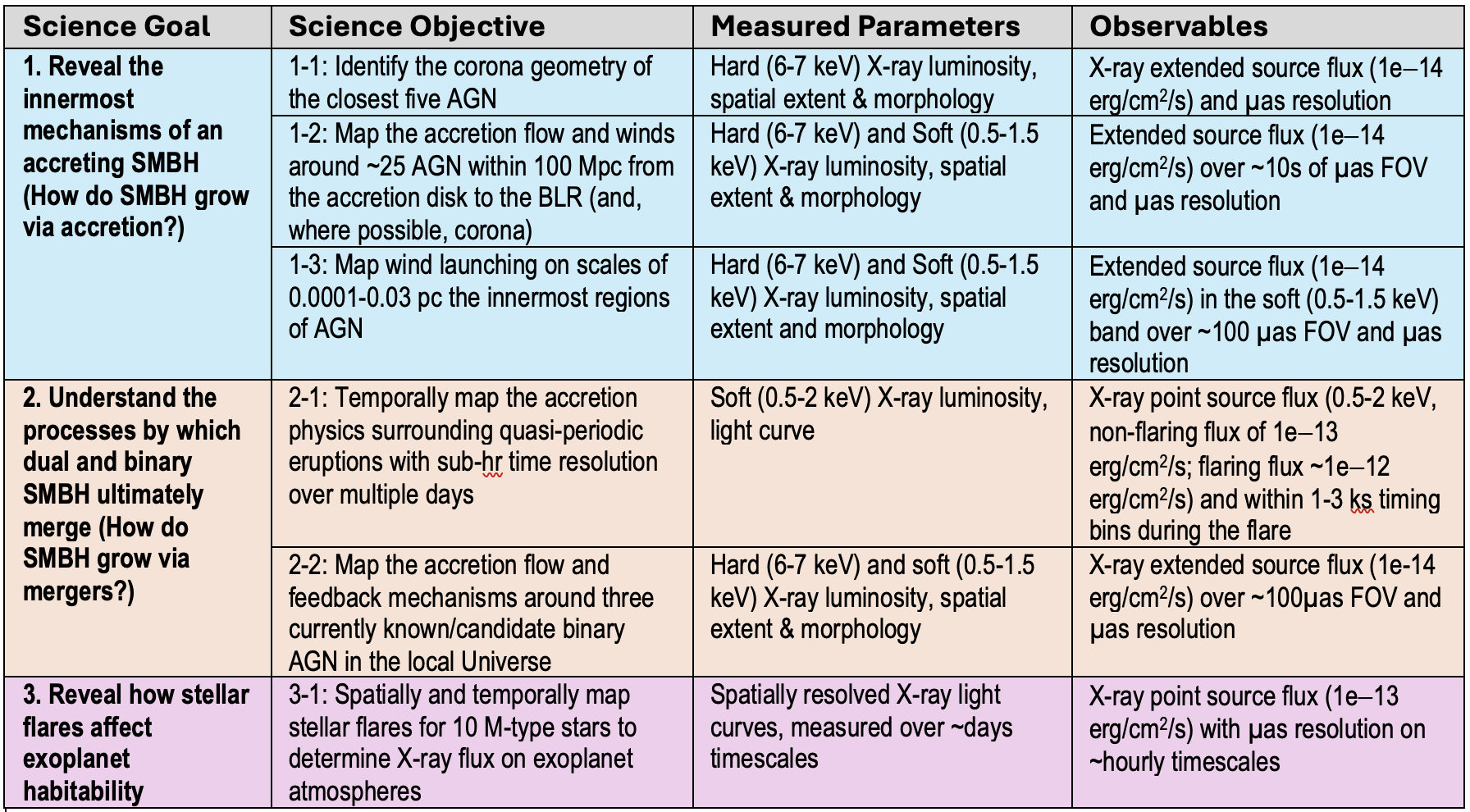}
\caption{\small \textbf{Science Traceability Matrix (STM)} developed through the NIAC Phase I process. This STM traces the science goals for three primary science cases: black hole accretion, the sources of gravitational wave events, and exo-space weather.}
\label{Fig:stm}
\end{figure}

{\bf Science Case 2: Dual and Binary AGN and Gravitational Wave precursors}. Accretion science also entails resolving dual SMBH in dual active galactic nuclei (AGNs) as gravitational wave (GW) precursors, and gaining a deeper understanding of other merger-driven mechanisms of black hole growth. Gravitationally bound binary AGNs with pc to sub-pc separations – which represent gravitational wave precursors – are currently only resolvable via radio Very-long-baseline interferometry (VLBI) and would therefore be prime targets for AEI.

{\bf Science Case 3: Stellar coronal physics and exoplanet habitability.}
For exoplanets, high-energy photons and energetic particles generated by stellar flares and coronal mass ejections interact with planetary atmospheres and impact plant habitability. Having spatially resolved imaging of stellar coronae would provide unprecedented insight into the effects of flares on exoplanetary systems. The X-ray ``flooding'' of exoplanet atmospheres by their parent stars is key to understanding habitability.

\subsection{Science Traceability Matrix}
\label{stm}

\begin{figure}[h]
    \centering
\includegraphics[width=0.52\textwidth]{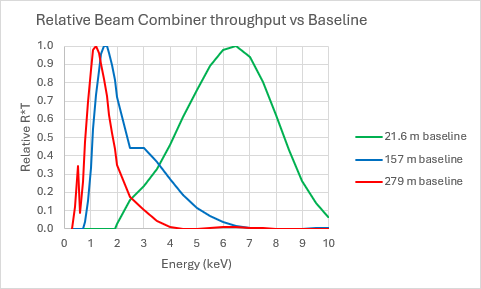}
\includegraphics[width=0.46\textwidth]{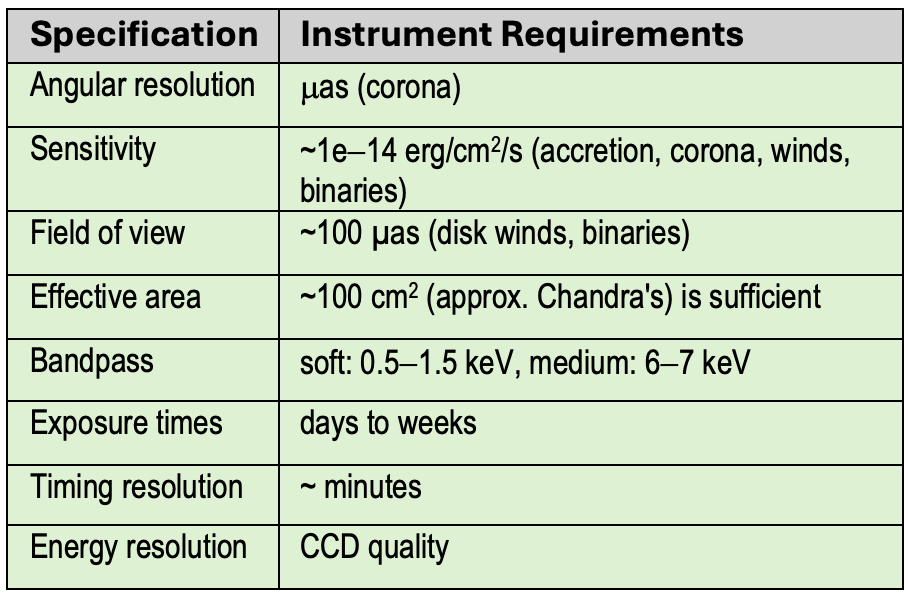}
\caption{\textbf{AEI Energy bands and technical specfications.} Left: Baselines were derived from the energy bands where we tune the interferometer for maximum science return according to the STM. This plot shows the width of the bands to which AEI is sensitive. Right: Technical specifications required to successfully complete the science objectives noted in the STM.
}
\label{Fig:energy_bands}
\end{figure}

Goal 1 of our study was to flow down the science requirements into mission design parameters for the primary $\mu$as interferometer (see Section \ref{combiner}) by constructing a preliminary STM (Figure \ref{Fig:stm}). The first aim was to determine energy bands, which place requirements on the baselines (see Figure \ref{Fig:energy_bands}). Second, we determined the target flux levels that would create effective area requirements and thus exposure times to provide the sensitivity that is needed for detailed SMBH corona and accretion disk studies at 1 $\mu$as (see Cann et al. in prep for more detail). Third, we sought to determine the area and exposure times required for stellar astronomy to study stellar flares, which impact planet formation and habitable worlds. The overall technical requirements necessary for successful completion of the science objectives can be found in Figure \ref{Fig:energy_bands}. The table compiles the most stringent technical specifications of each science case to identify ideal mission parameters. 

In this section, we discuss the technical specifications needed to successfully accomplish accretion science. In particular, we present the notional mission description created for the Mission Design Lab study, that served as a precursor to developing a science traceability matrix. The dual agn and stellar cases are presented in  Appendix A (\ref{AppendixA}). An even more thorough discussion of all of the key AEI these science cases, including the current state of the art and the specific science questions used to determine these parameters, can be found in Cann et al. 2026 (white paper in preparation).

\subsection{Notional Mission Description for SMBH and Accretion} 

The ability to spatially resolve the innermost regions of an accreting supermassive black hole in the X-rays would open up a transformative new discovery space to expand our understanding of these extreme objects: \textit{What is the geometry of the central engine of an AGN at the smallest scales? How does this geometry change as a function of fundamental parameters, such as black hole mass, accretion rate, spin, etc.? How do the winds that are driven from the disk shape and move the material feeding the SMBH itself? }


\begin{itemize}
\item[] \textbf{Angular resolution:} To observe in proximity to the AGN accretion disk, we need to resolve scales on the order of $1\times10^{-5}$ to $1\times10^{-6}$ pc in local AGN (corresponding to $1-10~\mu$as). Resolving the X-ray corona would require $\mu$as scale resolution. 

\item[] \textbf{Bandpass:} One band from $6-7$ keV to comprise the Fe K complex; at least one soft band within $0.5-1.5$ keV.

\item[] \textbf{Sensitivity:} At least $10^{-14}$ erg~cm$^{-2}$~s$^{-1}$ (lowest estimate: $1\times10^{-16}$~erg~cm$^{-2}$~s$^{-1}$) 
Ideally this would be per resolution element to obtain the best imaging quality, but this can be multiple resolution elements binned together.


\item[] \textbf{FOV}: $\sim100$ $\mu$as (or the ability to tile/mosaic).

\item[] \textbf{Spectral Resolution:} CCD quality is ideal, or rough multi-band imaging is sufficient.

\item[] \textbf{Observation exposure times:} Approximately a week or more.

\end{itemize}

\section{Note on Assumptions for the Primary Instrument}
\label{combiner}

\begin{figure}[h]
    \centering
\includegraphics[width=0.7\textwidth]{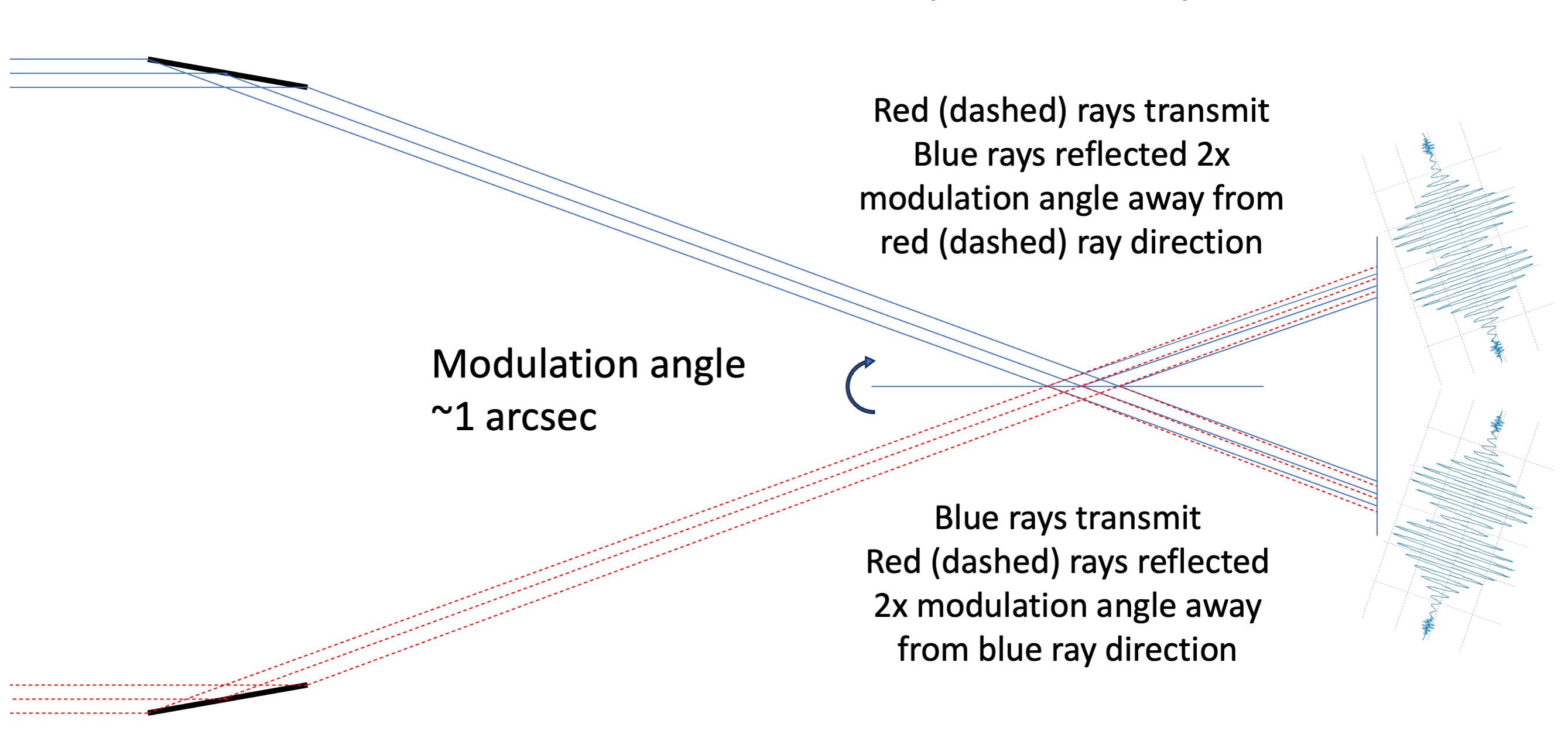}
    \caption{\small {\bf Thin film beam combiner optical layout.}}
    \label{fig:combiner}
\end{figure}

Our Phase I study was built around requirements for the {\it primary} instrument, which is still at a low technology readiness level (TRL). We are pursuing innovation via non-NIAC funds to revitalize an interferometry testbed and produce fringes. Our optical system design is based on a Michelson interferometer \citep{willingale2004}, where two X-ray beams are collected, separated by a baseline with a central wavelength $\lambda$ and brought together on a detector. The innovative element is a mirror beam splitter concept that eliminates long focal lengths, providing a configuration that works over a wide range of wavelengths in a single assembly (Figure \ref{fig:combiner}). 

In this design, about 30\% of each arm reflects from and transmits through the beam combiner, so both arms can be utilized fully, not just the reflected side of one arm and the transmitted side of the other. Rays from each pair of collecting mirrors land on separate regions of the detectors, separating the interference patterns. The competing spectral response of the reflection and transmission of the beam combiner also provides a filtering effect as a function of angle. For any incident angle, higher-energy X-rays are rejected in the reflected path by the critical angle of total external reflection while lower-energy X-rays are absorbed by the substrate in the transmitted arm. 

The challenge is to have thin (<1 $\mu$m) substrates that can maintain the < $\lambda$/50 figure quality for soft X-rays and survive the launch environment. The EUV lithography field has pursued very thin, optically transparent components for a decade \citep{gallagher2015,nam2022}, and there are off-the-shelf pellicles >1 $\mu$m that could be used for $6-7$ keV energies. We are currently working in the lab with materials to test our design.

X-ray beam splitters have not yet been used for astronomical applications, but do have other uses. Soft X-ray beam splitters are used for X-ray holography to enable the 3D imaging of material microstructures, and also for soft X-ray interferometry to probe laser-produced plasmas \citep{PhysRevLett.74.3991, Liu:08}. Bragg X-ray beam splitters have been developed for use in quantum X-ray optics \citep{powers2025}. X-ray beam splitters are also used for X-ray laser cavity output couplers, to steer and manipulate beams in laser systems for advancd laser research \citep{10.1117/12.964826}.

\section{AEI Mission Concept}
\label{mdl}

Our mission concept includes seven formation-flying spacecraft: one detector spacecraft (DS), and three collector spacecraft (CS) pairs (two identical spacecraft each). Each CS pair is separated by a specific baseline distance (Figure \ref{fig:constellation}). The set of primary mirrors on each CS pair collects X-ray photons and reflects them at grazing angles, directing them to a secondary optic on the DS (Figure \ref{fig:ds_optical_path}). Before assessing the formation flying, we created the optical design for the interferometer satellite array (Figure \ref{fig:constellation}) and set angular resolution requirements of $\sim0.1''$ and $\sim1$ mas for the supporting X-ray spotting telescopes. The original concept from our Phase I proposal is discussed in Appendix B (\ref{AppendixB}), which we have expanded upon.

\subsection{X-ray Optical Design}
\label{optics}

We studied three energy channels for the CS pairs based on mapping the X-ray continuum and key astrophysical emission lines for relevant science cases \citep[e.g.,][]{weaver2026}, requiring three sets of collector mirror spacecraft ($\sim0.8-1.5$ keV), ($\sim1.4-2.0$ keV), ($\sim5.8-7.0$ keV). See also the throughput curves in Figure \ref{Fig:energy_bands}. For $\mu$as resolution, the collecting mirror separation is D = $\lambda$ / 2x1 $\mu$as = 21.6 m for the inner mirror pair. The distance from the collecting to the combining optic is 2 km, and because of the nature of the beam combining optic (Section \ref{combiner}), it can be on the same spacecraft as the detector layout.

\begin{figure}
\centering
 \includegraphics[width=0.93\textwidth]{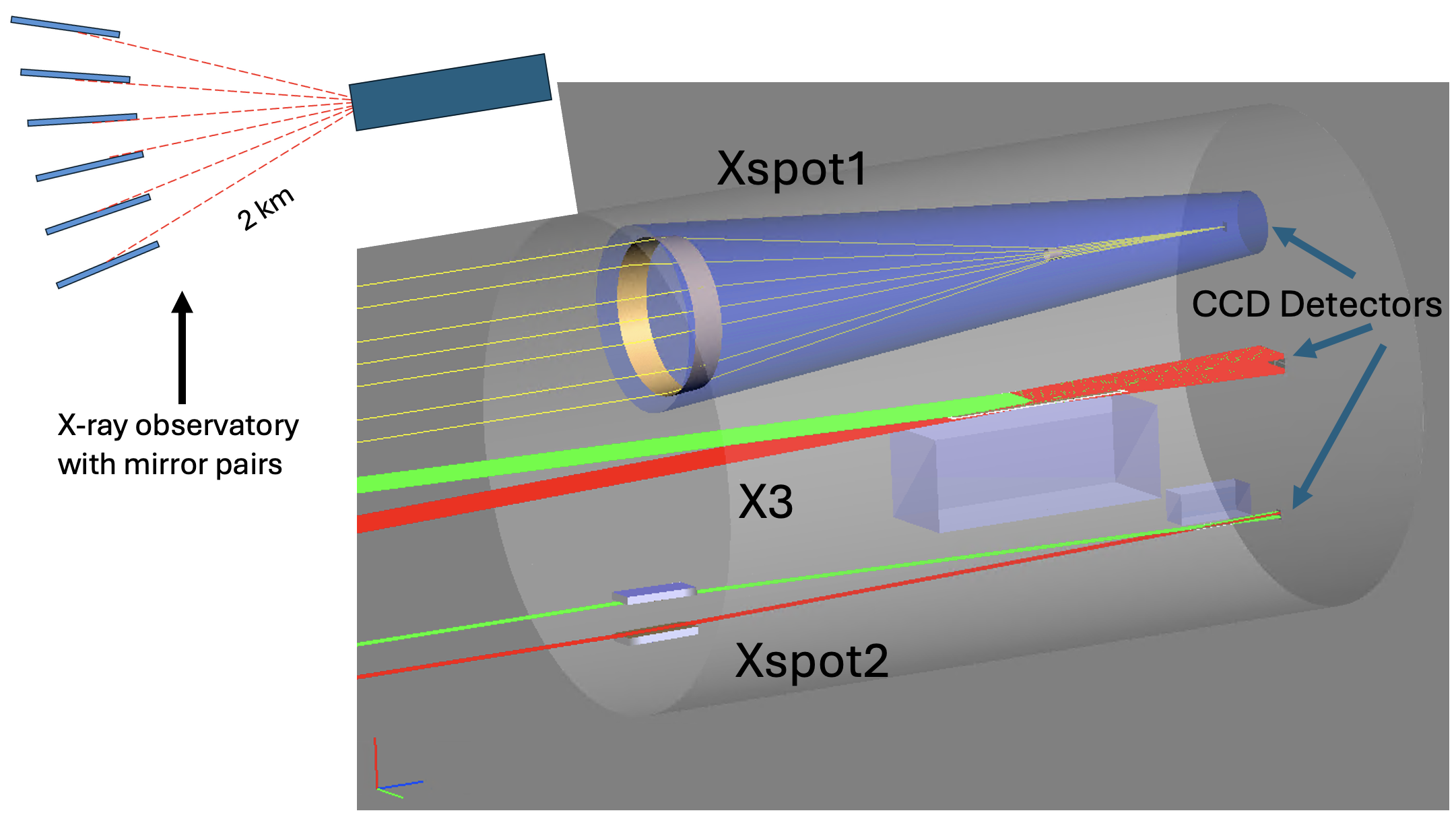}
    \caption{\small {\bf AEI optical paths.} From an optical design standpoint, this figure illustrates how the incoming X-rays are focused via all three X-ray telescopes onto the focal planes within the detector satellite housing. The transparent boxes represent figure error monitoring devices for the two beam splitters. Top inset: Adapted from W. Cash talk in 1999. For the observatory, X-rays reflect from flat mirrors held nearly parallel to the incoming light (from the left). The pairs of mirrors that form three baselines reflect the X-rays toward the detector satellite.
}
   \label{fig:ds_optical_path} 
\end{figure}

We created our design for the DS with a focus on the prime $\mu$as interferometer. The optical paths and relative placements of smaller telescopes to achieve the required target acquisition of the prime interferometer are shown in Figure \ref{fig:ds_optical_path}. We note that the smaller embedded telescopes are not meant to be unique solutions. Besides off-the-shelf optical star trackers for pointing, we require two X-ray spotting telescopes. The X-ray spotting scope 1 (Xspot1), shown in Figure \ref{fig:ds_optical_path} as a Wolter I mirror, requires $<1''$ resolution (Chandra-like resolution or better). The second X-ray spotting scope (Xspot2) needs to reach $\sim$mas resolution. For simplicity we envision here a beamsplitter similar to the primary optic of the $\mu$as interferometer, labeled X3. Both the Xspot2 and X3 beamsplitter assemblies would require an accompanying optical figure monitoring system. 

We calculated effective areas and the necessary field of view (FoV) to achieve the target acquisition and centering.\ To operate with resolving powers that are several orders of magnitude apart, the angular FoV of each successively more powerful telescope must be at least as large as the point spread function (PSF) of the less powerful telescope. 

The {\bf Xspot1} telescope has to locate the X-ray target and provide adequate collecting power. There are several solutions for Xspot1, but for our purposes and to fit within the spacecraft volume, we examined two options: (a) a  highly polished Wolter I design Chandra-type inner mirror shell (0.65 meter diameter) to reach a PSF similar to Chandra's $\sim0.5''$, and (b) a Wolter II design that achieves better resolution.

\begin{wrapfigure}{R}{0.4\textwidth}
\centering
 \includegraphics[width=0.39\textwidth]{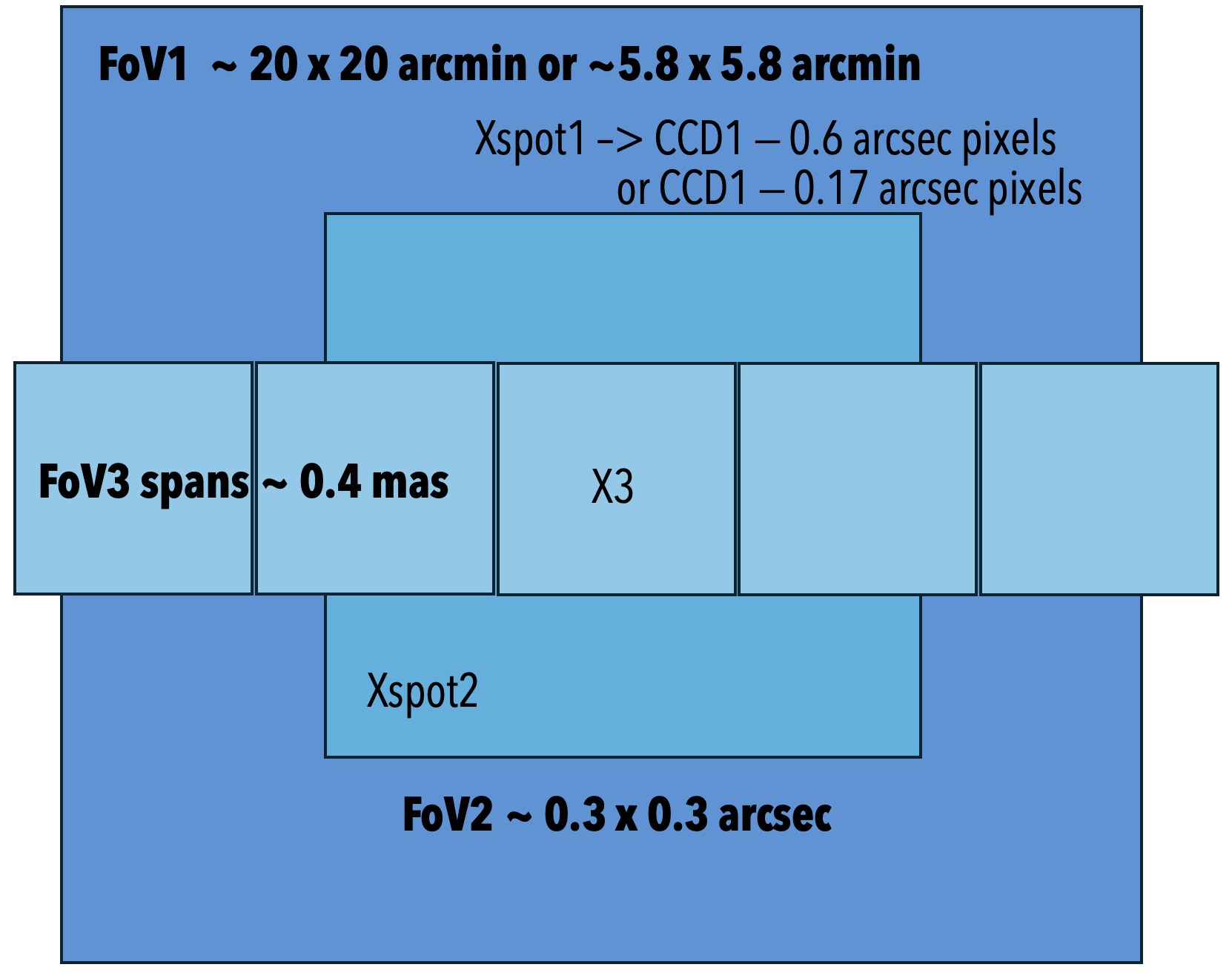}
    \caption{{\bf Instrument field of view overlay for comparison.} Relative comparison; not to scale.
}
   \label{fig:fov} 
\end{wrapfigure}

We envisioned having a $\sim0.1-0.2''/$px plate scale for Xspot1, but with only a 5 meter focal length available (Section \ref{spacecraft}), that is not possible with the Wolter I design. The plate scale is given by PS = (pixel size ($\mu$m)/telescope focal length (mm)) x 206. Assuming large format X-ray CCDs with 15 $\mu$m pixels, we derive PS $\sim0.6''$/pixel. For a 2048 x 2048 pixel CCD, this provides a $20'\times20'$ FoV (Figure \ref{fig:fov}).

A Wolter II grazing incidence mirror design uses a paraboloid and inner hyperboloid mirror to achieve a longer effective focal length and better resolution. This design is not traditionally used for X-ray astronomy. But as a narrow field imager, this could be an option for a spotting telescope.
A Wolter II mirror has been successfully flown as one of the payload instruments on the Extreme Ultraviolet Explorer \citep[EUVE; ][]{malina1980}.

\begin{wrapfigure}{R}{0.4\textwidth}
\centering
 \includegraphics[width=0.39\textwidth]{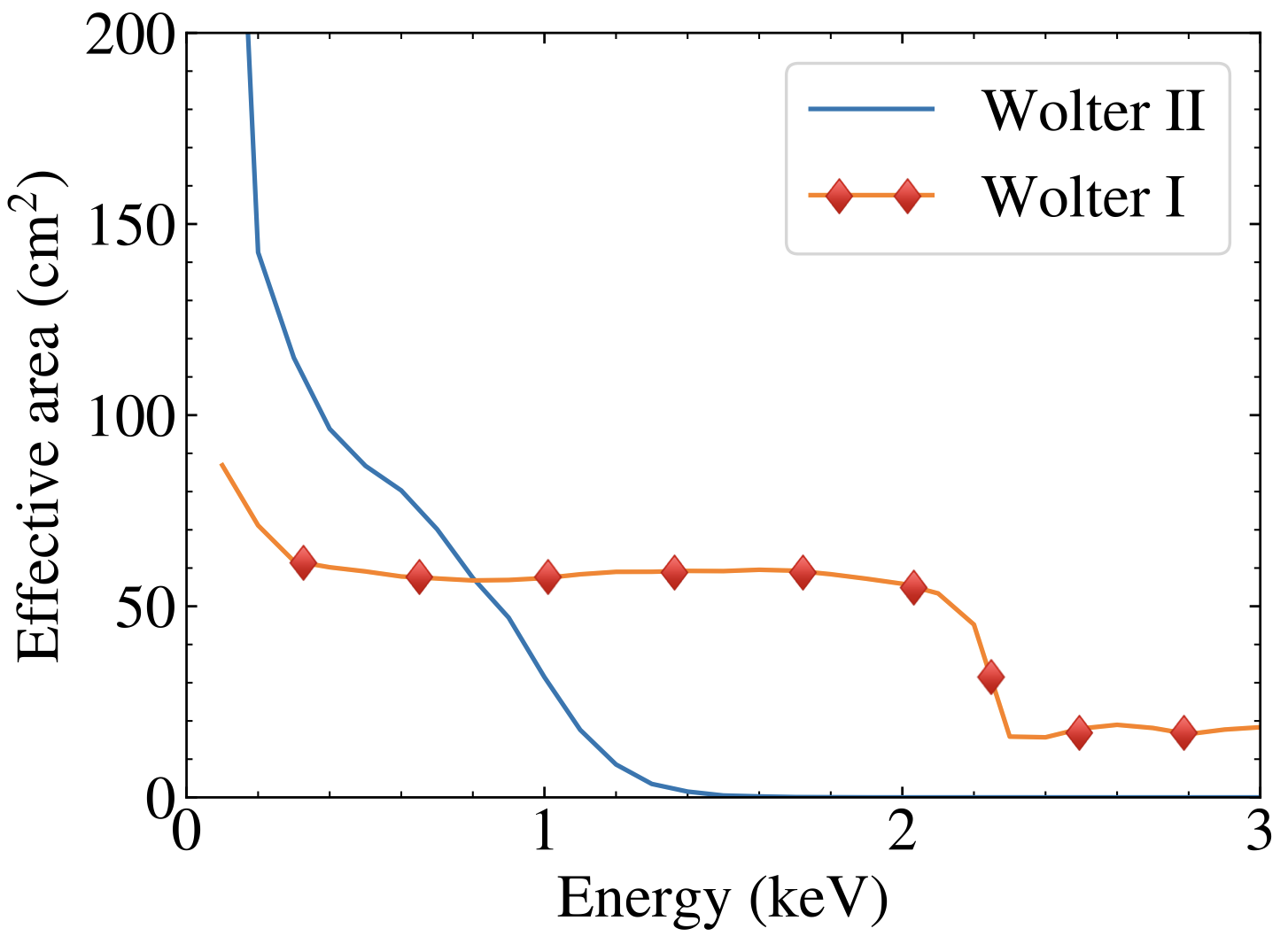}
    \caption{{\bf Xspot1 effectice area comparison.} Blocking filter is not included. 
}
   \label{fig:area} 
\end{wrapfigure}

A Wolter-II design with a single 1-meter diameter, 200 mm long collecting mirror, and a length of 4 meters, provides an effective focal length of 16.34 meters. Assuming 2048 x 2048 pixels and a 13.5 $\mu$m pixel detector, the PS = $0.17''$/pixel and the FoV is $5.8\times5.8$ arcmin (Figure \ref{fig:fov}). We calculated the effective collecting area for both designs, (Figure \ref{fig:area}). For a spotting telescope to locate the bright celestial targets we would study with AEI, an area of approximately 60 cm$^{2}$ at $\sim1$ keV is enough for target acquisition. While this area is adequate, more study is needed to optimize Xspot1, especially if we require a spotting telescope at energies $>2$ keV.

The {\bf Xspot2} uses a single pair of 0.5 meter long collector mirrors with a baseline of 210 mm that provides a resolution of 0.33 mas at 1.51 keV (Al K$\alpha$) and 0.16 mas at 6.5 keV (Fe K$\alpha$). The pellicle is 0.5 meters long and the detector is 500 mm downstream of the center of the pellicle. The total length is 4.5 meters. Modeling a single point source provides a full FoV of $\sim0.3\times3$ arcsec (Figure \ref{fig:fov}). We could potentially double the FoV with 1-meter long mirrors. 

Using the entire width of a 60 x 60 mm CCD, the total collecting area at 6.5 keV is 6 cm (CCD width) x 50 cm (mirror length) x sin(0.75) = 3.9 cm$^2$. The reflectance of the collector mirrors is $80-90\%$ and the throughput of the pellicle is about $40\%$ at 6.5 keV. The quantum efficiency (QE) of the detector should be $95\%$ for these energies, so the total effective area becomes 3.9 cm$^2$ * 0.8 * 0.4 * 0.95 = 1.2 cm$^2$. 

Ideally, for Xspot2 to pick up and resolve targets from Xspot1, we want an area at least that of Chandra's ($\sim100$ cm$^2$ at $\sim6$ keV). However, we have not performed any trades and the beamsplitter assumption was only to meet our requirement of mas resolution for our DS volume. We can improve the area by increasing the size of the collector mirrors and adding CCDs. But we have no dedicated solution for Xspot2 and such a telescope requires detailed further study (Section \ref{future}). 

Alternate designs for Xspot2 could include diffractive/refractive elements such as refractive beam combiners \citep{2009ExA....27...61S}, or phase fresnel lenses to achieve near diffraction limited imaging \citep{RANKIN20254977}. There have been other designs suggested and current work proposed or ongoing (see Table 2 in \cite{weaver2026}. But in any case, just like $\mu$as resolution, this ability to achieve mas resolution is a key enabling technology for AEI.

For {\bf X3} we also calculated the angular resolution, FoV, and effective area. The X3 pellicle is 1.2 meters long and 400 mm wide to cover the depth into the page for 5 CCDs that are $\sim60\times60$ mm each (Figures \ref{fig:ds_optical_path} and \ref{fig:fov}). At 1 keV, using 2-meter long and 420 mm wide collector mirrors and a beamsplitter modulation angle of 1.0 arcsec, the angular resolution is $\sim0.9$ $\mu$as. For a 13 $\mu$m pixel CCD, the X3 instrument FoV spans $\sim0.4$ mas (Figure \ref{fig:fov}). 

The effective area for X3 is driven by the CCD area and the mirror width, assuming a fixed mirror length. Using the group of five 60 x 60 mm CCDs with the same mirror, pellicle throughput, and QE, the effective area is $\sim32$ cm$^2$ at 1.5 keV and $\sim19$ cm$^2$ at 6.5 keV. Our original design had two large collector mirror flats in the 6.5 keV channel and we simplified this for our MDL study (see below), but this would have automatically doubled our 6.5 keV area to $\sim38$ cm$^2$. Also, for the widely separated energies from the widest and narrowest baselines (green and red curves in Figure 6) we can use the same detector pixels for both channels, i.e., the 279 meter channel and the 21.6 meter channel -- using the energy sensitivity of the detector to discriminate which channel is contributing. This could triple or quadruple the effective area for each baseline. 

 We can also improve the area by adding another row of CCDs or making the collector mirror longer, which would fit in our spacecraft design (there is room - see Figure \ref{fig:collector}). Given our goal to examine an observation sequence for a feasible area, we assumed for this study an area of 100 cm$^2$, which is about the same as Chandra's at $\sim$6.5 keV. For simplicity, we also assumed 100 cm$^2$ for low energies (stellar science) for our concept of operations study (Section \ref{operations})

\subsection{Mission Design Lab (MDL) Study and Scope}
\label{mdl}

Our MDL objective was to increase the Concept Maturity Level (CML) for AEI from CML 1 to 2 or 2+ to demonstrate feasibility. For simplicity, all six CS were assumed to be identical. A top level summary of input and study adjustments is shown in Table 1.

\begin{table}[h]
    \centering
    \caption{  } 
    \begin{footnotesize}
\begin{tabular}{|| l | c | c ||}
    \hline
    \multicolumn{3}{|c|}{{\bf Top Level Input and Study Adjustments}} \\
    \hline\hline
Launch Readiness Date     & 2040s &  \\
Mission Life & 5 years (goal)  & 2 years (this study)  \\
Orbit &  L2 Halo &     \\
Mission Class  & C &   \\
Launch Vehicle & big &   \\
Satellites & 7 & all free flying  \\
Observations & long stares & non-continuous  \\
\hline
   \multicolumn{3}{|c|}{Derived and compared with Chandra} \\
   \hline
Total launch Mass &  9,700 kg  & 5,860 kg   \\
Electrical load & 2452 Watts (DS) & 2350 Watts \\
 & 2180 Watts (per CS) & \\
Total Volume & $\sim90$ m$^3$ & $\sim30$ m$^3$ (approx.)\\
\hline
    \end{tabular}
    \end{footnotesize}
    \label{tab:input}
\end{table}

Our goal for the mission lifetime is five years. However, as we progressed through the MDL study, it became clear that observations with this architecture that hold such a precise pointing are ``expensive'' and require resources. Specifically, the $\Delta$$v$ budget only accommodates 50 targets based on our preliminary exposure times (see Section \ref{operations}, Table \ref{tab:deltav}). The $\Delta$$v$ budget is defined as the total change in velocity needed for the spacecraft to perform the sum total of all planned maneuvers during its mission. The $\Delta$$v$ budget also represents the fuel budget and so enough fuel must be carried on board to execute all planned maneuvers. For our target list (Table \ref{tab:deltav}), the total minimum and maximum $\Delta$$v$ is 399 and 638 m s$^{-1}$, respectively. This is large for a space mission lifetime. Such a large $\Delta$$v$ requires a large reserve of propulsion, which suggests that a two year lifetime for this design is more likely than five. However, we have not yet optimized the target planning. We are thus keeping five years as a goal with the assumptions that (1) there could be development of technology in the area of propulsion to accommodate a longer mission and (2) observation planning tools will be developed to more properly model this design (Section \ref{technology}).

Based on our STM, we require a FoV of at least 100 $\mu$as, which we can meet in a single exposure without tiling (Figure \ref{fig:fov}). For the effective area of approximately Chandra's, we require exposure times between 3 and 19 days, Table \ref{tab:deltav}). The most stringent precision-pointing requirement for the telescope is driven by maintaining fringes on the X3 detectors. As illustrated in Figure \ref{fig:constellation}, the formation flying array requires attaining and maintaining pathlengths to within 100 picometers. Other requirements are accuracy: $0.5''$ (outer two mirror pairs), $0.1''$ (inner pair); and knowledge $0.05''$ (outer two mirror pairs), $0.01''$ (inner pair). The X-ray mirrors in the CS need to be maintained to $\sim$mas stability (tip/tilt) during observations.

 We assume that the six CS will operate at room temperature with the science boresight precision maintained using two LISA-type laser units on each spacecraft that point toward the DS and their baseline counterpart (maintaining a laser ranging triangle - Figure \ref{fig:constellation}). We initially assumed that we would need the full range of the LISA lasers to maintain the stability along the line-of-sight to the target (picometer stability), but we were able to relax this requirement to 100 picometers. We assume there can be no gimbal articulation during an observation, but due to natural drift, we chose to break up daily observations with a one hour “maintenance” period per day driven by the need to regularly re-orient the solar arrays (Section \ref{operations}). 

\subsection{Mechanical Systems}
\label{mechanical}
Here we describe the mechanical aspect of the architecture, such as the spacecraft details, instrument packaging, and future launch vehicle possibilities.

\subsubsection{Note on Telescopes and our Modular Approach}
\label{modular}

\begin{figure}
\centering
 \includegraphics[width=0.8\textwidth]{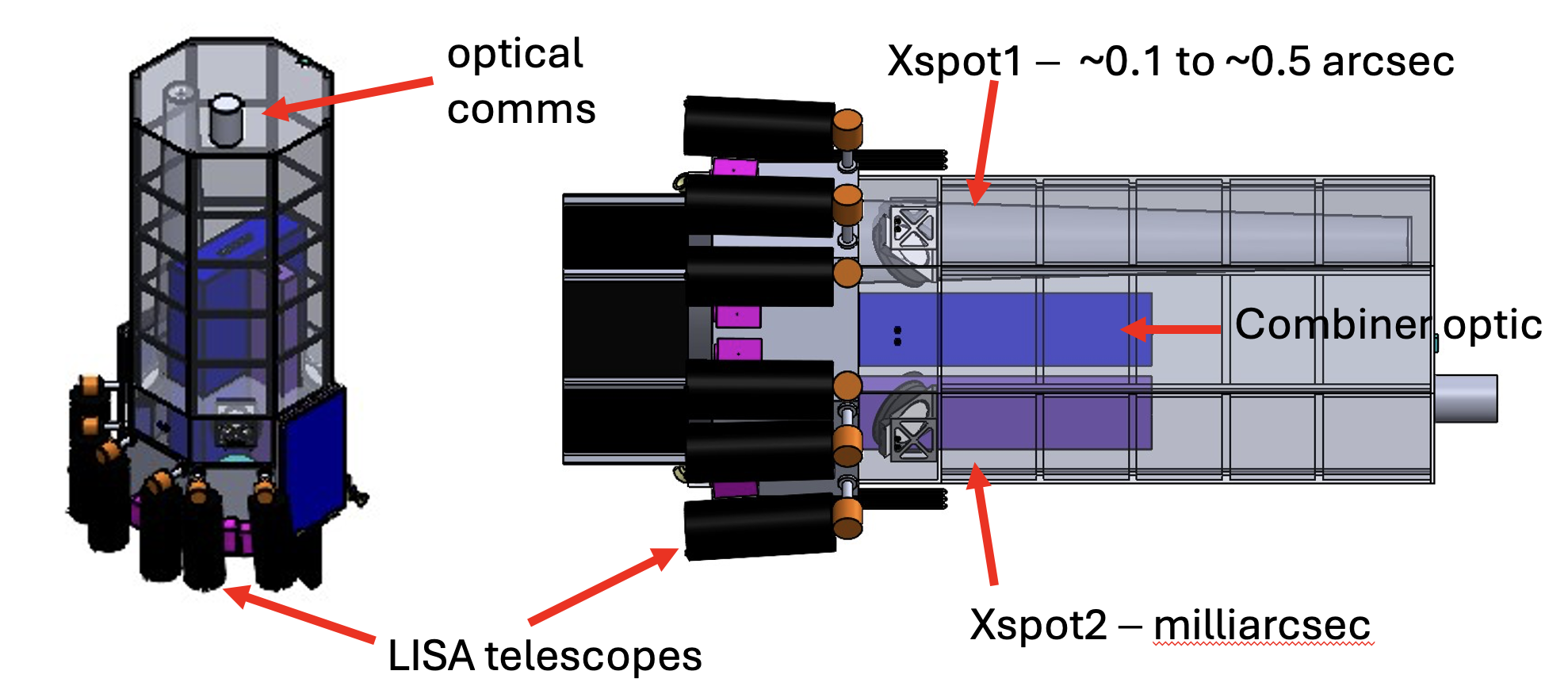}
    \caption{\small {\bf Modular approach for the AEI detector spacecraft.}
}
   \label{fig:modular} 
\end{figure}

As discussed in Section \ref{optics}, our DS design includes multiple smaller (embedded) telescopes to achieve the target acquisition and centering for the prime $\mu$as interferometer (Figure \ref{fig:modular}). There are three types of finder scopes. The first are off-the-shelf optical star trackers for $\sim1''$ location, which will be used to orient the array based on optical star fields. Then there are two progressively stronger X-ray spotting telescopes, Xspot1 and Xspot2. Finally we have the large beamsplitter assembly that provides $\mu$as resolution (labeled as the combiner optic in Figure \ref{fig:modular}). We include the LISA telescope pairs mounted on the DS and pointed toward the six CS and along the target line-of-sight (science boresight) to attain and maintain pathlengths at 100 picometer knowledge. 

The innovation of our design is that we have a modular system with three separate X-ray optics, all perhaps different, packaged in the same telescope, but designed to be interchanged and re-examined as new technologies develop. The only true requirement is using the X-ray spotting telescopes at coarser pointing to align and center to the $\mu$as scales required. We have chosen the modular approach to maintain flexibility and longevity for this architectural design. We do not know how the technology development for X-ray optics will evolve over the next few decades and since flagship missions take decades to mature, we want to remain as flexible as possible for all of the X-ray telescopes. A modular approach is also more flexible to the architecture being open to adding new instruments or swapping out others. Our X-ray spotting telescopes can be {\it any future telescope} that gets to the needed angular resolution and can be packaged within a similar design.

\subsubsection{Innovation: LISA Telescope adaptation}
\label{lisa}

The telescopes designed for the LISA mission provide a crucial inter-spacecraft link to precisely maintain the distances between the LISA spacecraft used to measure gravitational waves \citep{2025CQGra..42w5007B}. 
These telescopes work by simultaneously transmitting and receiving coherent near infrared laser light (1064 nm), and this maintains the LISA formation stability to within picometers.


We have assumed that the LISA telescopes can be redesigned to meet the AEI detector/collector alignment requirements and packaged within/on the spacecraft. The LISA telescopes are Low Critical Technology Element (CTE) telescopes. They are well understood by both GSFC and industry, but at this time require a fair amount of Non-Recurring Engineering (NRE). NRE refers to the one-time, upfront costs that are associated with researching, designing, developing, testing, and qualifying a new satellite component or subsystem before the first unit is manufactured. The AEI distances are significantly smaller than LISA, so a lower power telescope could be developed to minimize the power resource need. One possible alternative is the Laser-Ranging Interferometer (LRI) on GRACE-FO, which tracks a 220 km distance to nm precision \citep{PhysRevLett.123.031101}. Another is the Compact Astrometric Alignment Sensor (CAAS) that reaches 30 mas resolution relative to intertial space (https://techport.nasa.gov/projects/146773).

\subsubsection{Spacecraft}
\label{spacecraft}

\begin{figure}
\centering
 \includegraphics[width=0.8\textwidth]{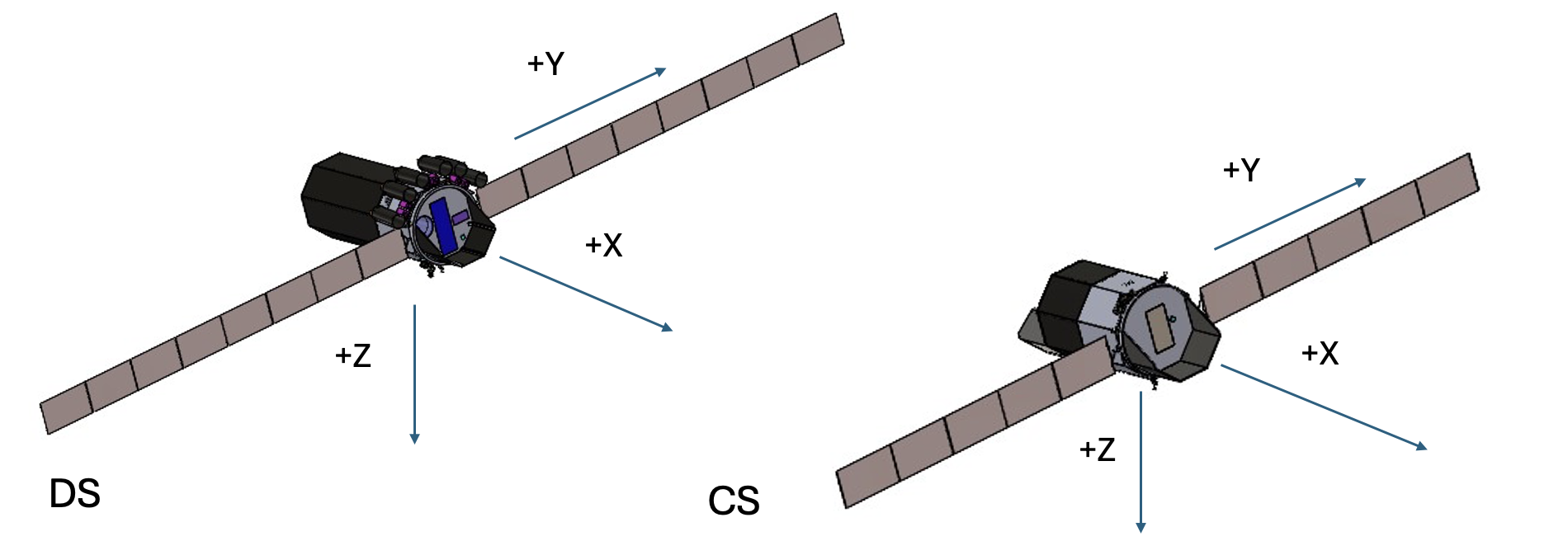}
    \caption{\small {\bf AEI Spacecraft types deployed}
}
   \label{fig:two_craft} 
\end{figure}

The driving requirements for our spacecraft (Figure \ref{fig:two_craft}) are (1) millikelvin (mK) stability of the DS optics and mirrors, (2) a $\mu$as alignment of the six CS, and (3) their specific payload requirements. For the general bus structure we assume the collector and detector spacecraft types can be manufactured from aluminum. All spacecraft will accommodate propulsion, avionics, communications, solar arrays, attitude control system components, and all necessary thermal hardware. The bus structure is assumed to have hinged closeout panels on the sides to access the harness and avionics, etc. The DS carries hydrazine propellant tanks with 180 kg of propulsion using monopropellant thrusters (engines that produce thrust with a single liquid propellant). The CS each carry hydrazine propellant tanks with 50 kg of propellant, along with similar monopropellent thrusters. The six CS also carry Hall effect microthrusters with 119 kg of Xenon. Hall effect thrusters differ from monopropellent thrusters by using electricity to ionize the propellant. A top level spacecraft comparison is listed in Table \ref{tab:tech}.

\begin{figure}[h]
\centering
 \includegraphics[width=0.85\textwidth]{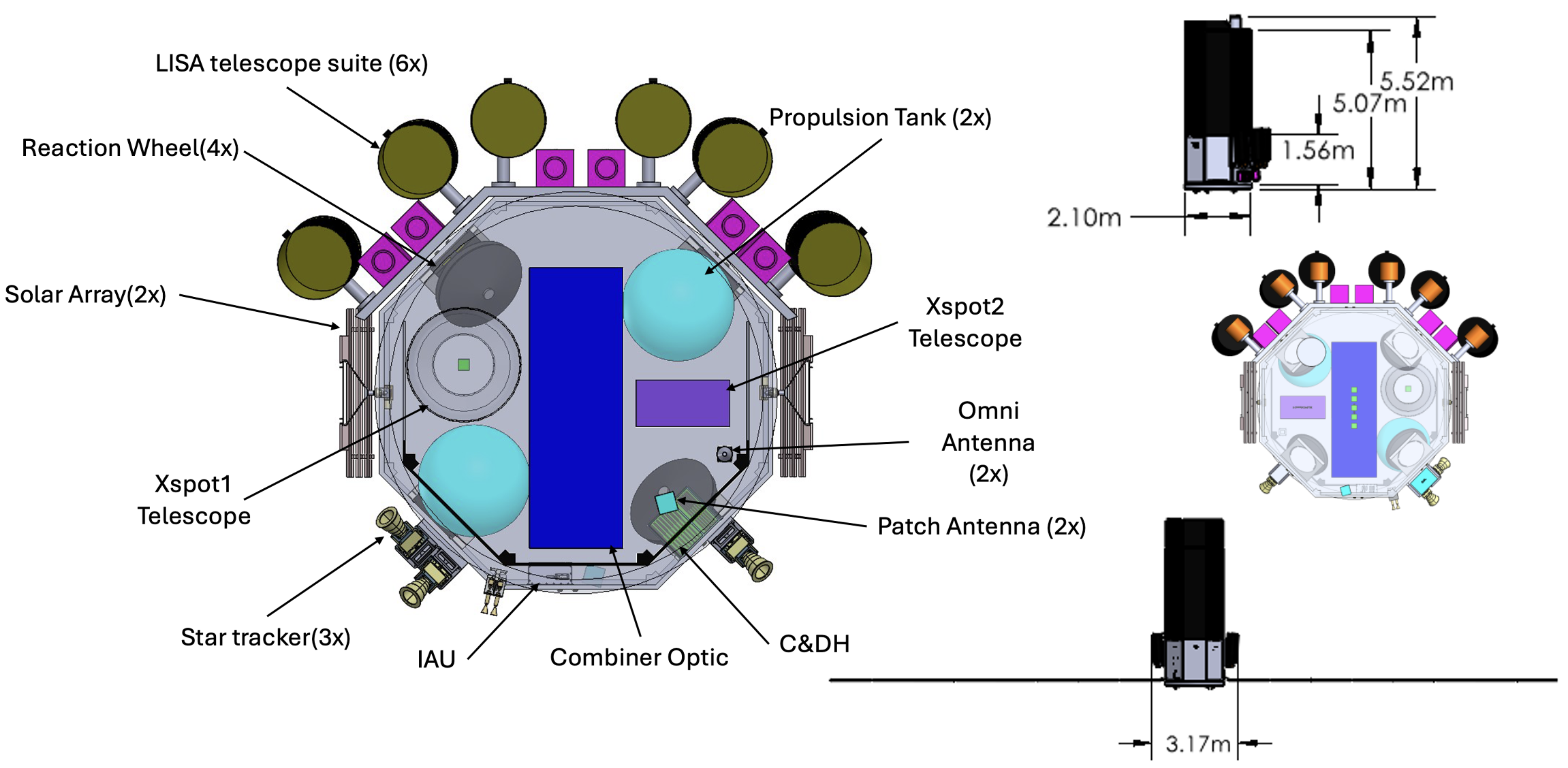}
    \caption{{\bf Detector spacecraft (DS)}. Large image is looking down the entrance aperture of the spacecraft, that points toward the target and the CS pairs. Smaller images are (top) stowed, (middle) transparent image of the detector end, and (bottom) deployed. IAU is the Integrated Avionics Unit.
}
   \label{fig:det_face} 
\end{figure}

The DS contains the beamsplitter assembly, its associated five-CCD array and figure monitoring interferometer, three LISA telescope / laser pairs, two X-ray spotting telescopes, and three optical star trackers. Various views of the spacecraft are shown in Figure \ref{fig:det_face}. It is 5.2 m long and 3.17 m wide near the end where the LISA telescopes are mounted. Figure \ref{fig:det_face} also shows the components and packaging of the optical elements. The beamsplitter assembly with its CCDs are positioned in the center of the volume with Xspot1 on the left and Xspot2 on the right. The mounting of the six LISA instrument suites are shown across the top. The detectors for the X-ray spotting telescopes are single, large-format CCDs. 

\begin{figure}
\centering
 \includegraphics[width=0.9\textwidth]{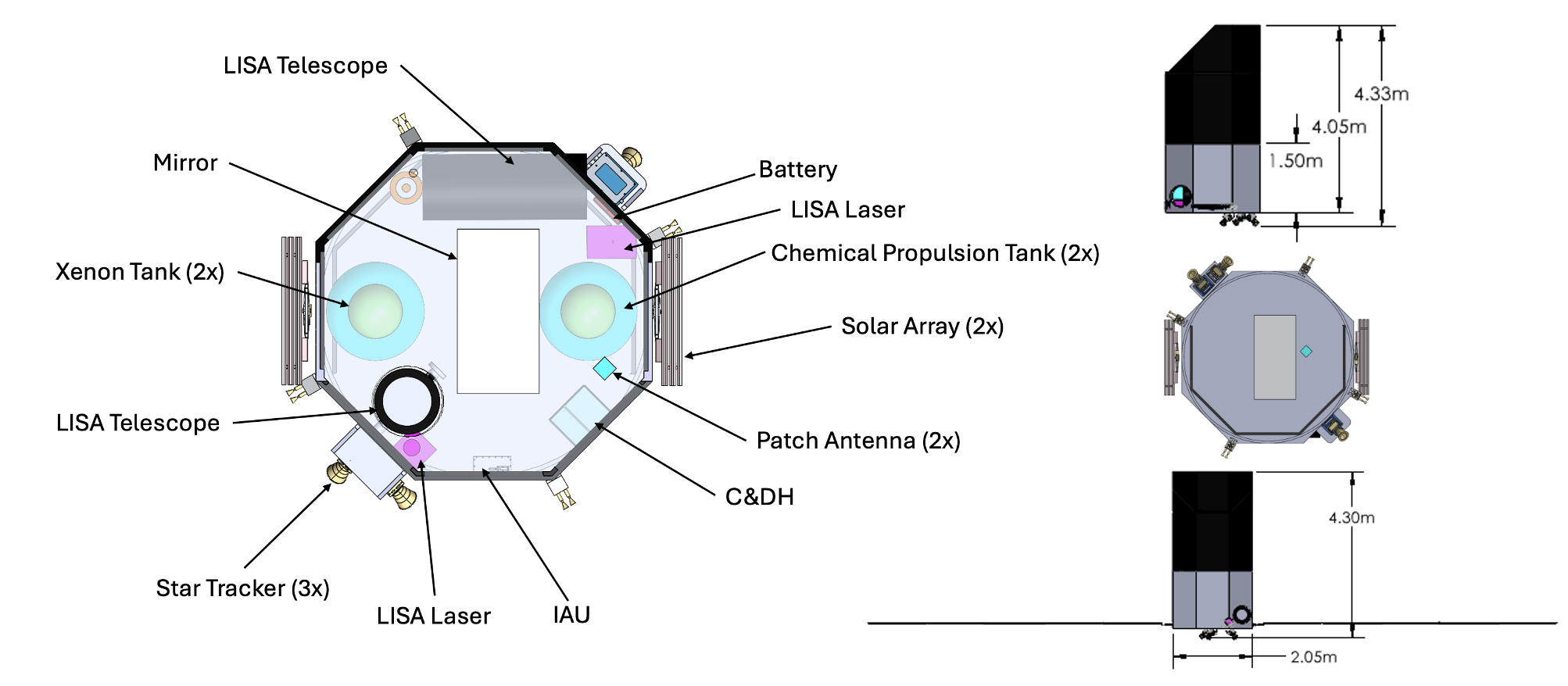}
    \caption{{\bf Collector spacecraft (CS).} The large image is a transparent view looking at the exit aperture end of the spacecraft, that points toward the detector spacecraft. The smaller images are (top) stowed, (middle) the entrance aperture that points to the target, and (bottom) deployed. The sunshade on the side that points to the target is not shown in these diagrams.
}
   \label{fig:collector} 
\end{figure}

Figure \ref{fig:collector} shows different views of a single CS. Each spacecraft contains a $2$~m$~\times~420$~mm mirror flat with active control to reflect X-rays at grazing angles, a LISA telescope / laser pair, and optical star trackers. The bus is 4.33 m long and 2.05 m wide near the top where the star trackers are mounted. The figure shows the side of the spacecraft that is facing the detector spacecraft with the LISA telescopes and lasers marked. Note the position of the LISA telescope toward the bottom left, which is mounted so as to point toward the detector craft, and the telescope at the back (front end) that is mounted to point toward the other CS that forms its baseline pair. For this study we assume the reflecting surfaces in the telescopes are close to perfect and we get laser ranging from the LISA system.

\begin{table}[h!]
    \centering
    \caption{  } 
\begin{tabular}{|| l | c | c | c ||}
    \hline
    \multicolumn{4}{|c|}{{\bf Spacecraft Comparison}} \\
    \hline\hline
    Spacecraft & Solar array area & Communications & Propulsion \\
     & per wing & & \\
    \hline\hline
 DS & 8.77 m$^2$  & S-band and optical & monopropellent  \\
 CS (x6) & 6.8 m$^2$ & S-band &  Hall-effect and monoprop  \\
\hline
    \end{tabular}
    \label{tab:tech}
\end{table}

\subsubsection{Solar Arrays, Electrical Design, and Operational Zones}
\label{zones}

Each spacecraft has deployable solar arrays (SA) and a peak operating temperature of 110 degrees C was assumed for SA sizing. We used a solar cell currently in full production at TRL 9 (cell area 69.31 cm$^2$). The SAs are modeled as two two-axis gimballed wings with a packing factor of $85\%$.  The DS SA total area is 15.84 m$^2$; the CS SA total area is 13.6 m$^2$. Each wing deploys directly from its respective bus and a two-axis gimbal will rotate each wing after deployment.  Because of the tight pointing alignment requirement, the design requires jitter dampers on the SAs to maintain stability. Deployable covers are required to protect the bore from sunlight in the case of off pointing or glint.

Our DS average electrical load is 2,452 W and the CS average load is 2,180 W. For comparison, Chandra operates at 2,350 W (Table \ref{tab:input}). The electrical design assumes a Direct Energy Transfer power system, which connects the SAs to the battery and the electrical bus without needing intermediate converters. The SAs track the sun up to a $20\degree$ off-pointing angle where pointing loss occurs, thus reducing the power output because the solar panels are not perfectly perpendicular to the sun. Nominal operations will use SA power and not the battery. 

\begin{figure}[h]
\centering
 \includegraphics[width=0.6\textwidth]{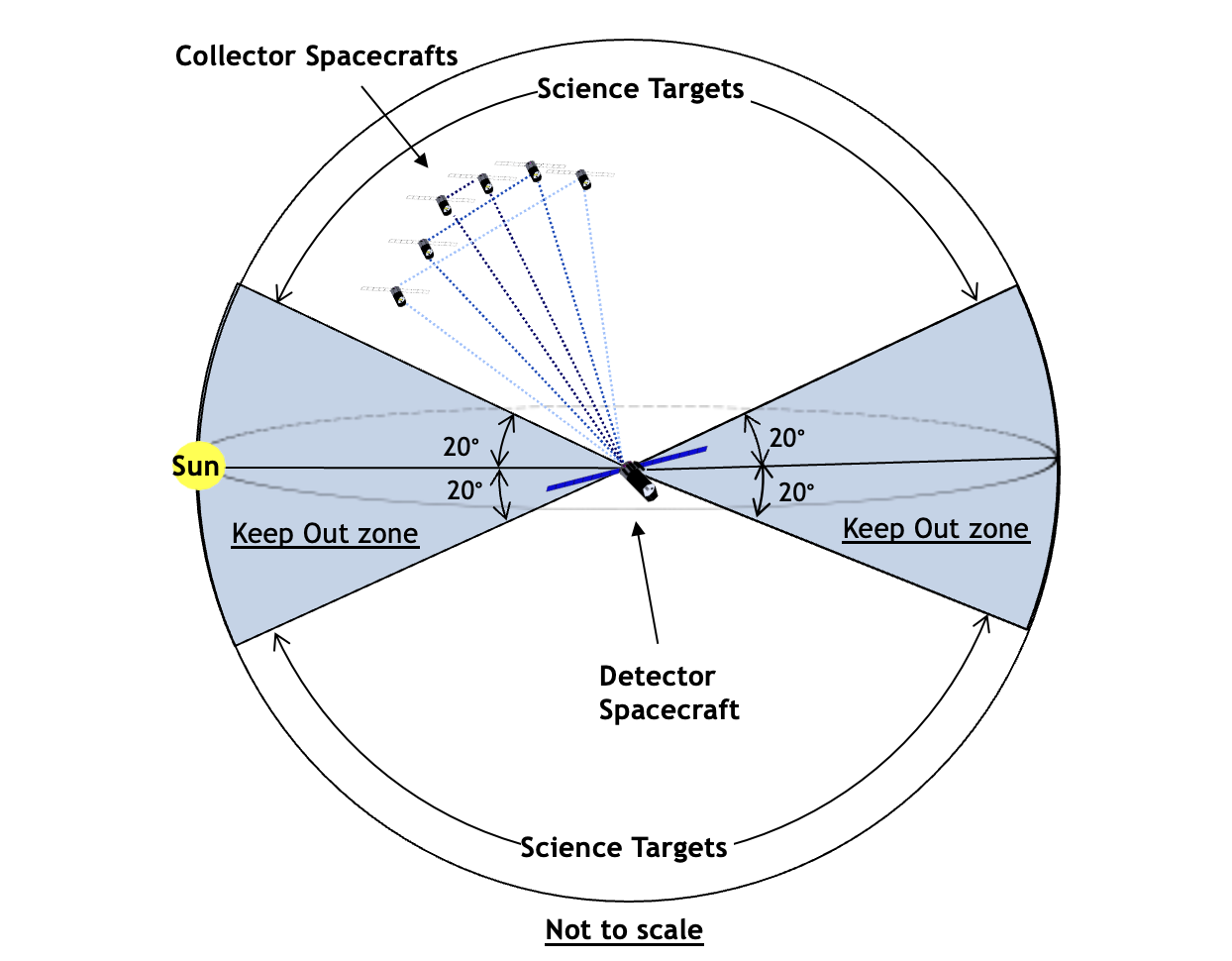}
    \caption{\small {\bf AEI Operational keep out zones.}
}
   \label{fig:avoidance} 
\end{figure}

Because of the unique situation that X-rays need to pass through the six CS, they are open on both ends.  This means that the back end of each CS is open to sunlight.  Some of this can be handled with sunshades, but this creates an “avoidance zone” for the array operation of $\pm20\degree$ from the ecliptic (Figure \ref{fig:avoidance}). The avoidance zone of $\pm20\degree$ does place a constraint on where we can look in the sky and thus on our potential target list. Celestial targets in that zone are not accessible with AEI. The operational plan would be to spend 1/2 of the year looking above the ecliptic and 1/2 of the year below. 

\subsubsection{Launch Vehicles}

\begin{figure}
\centering
 \includegraphics[width=0.9\textwidth]{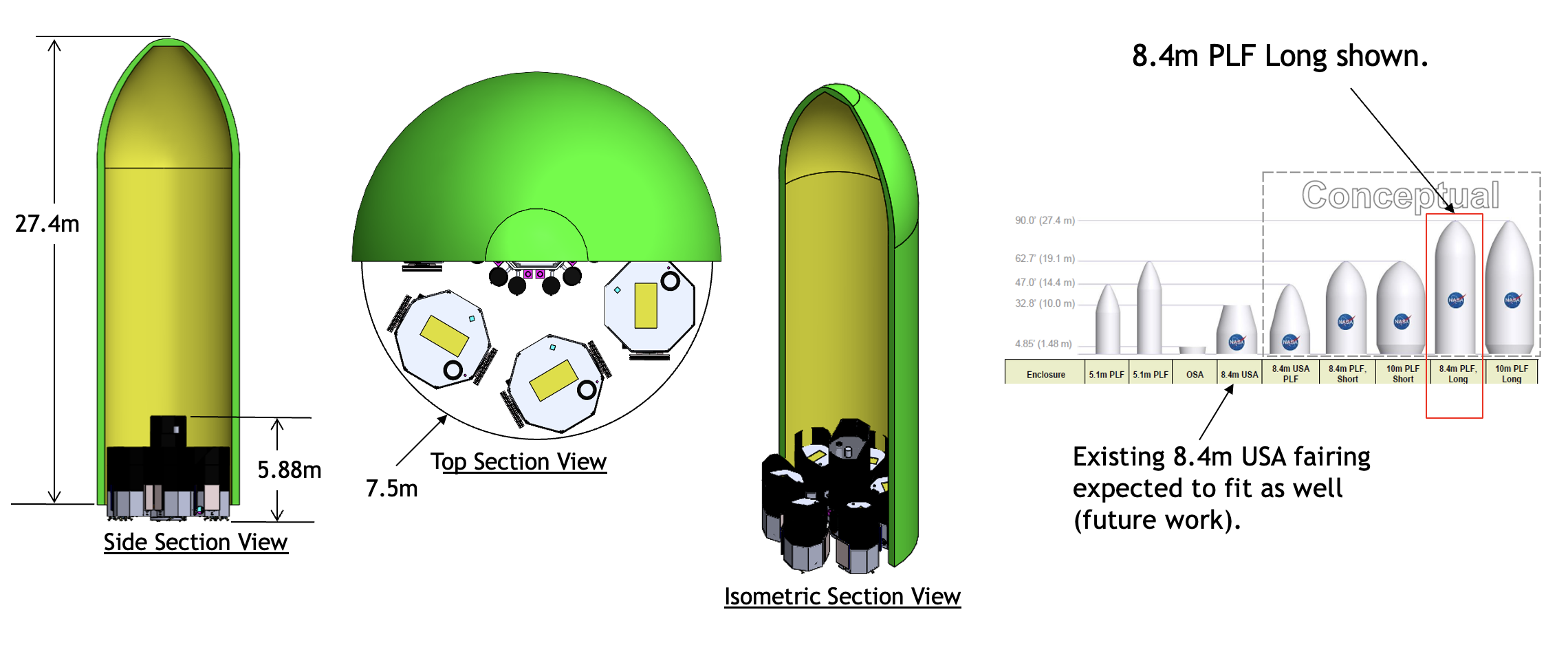}
    \caption{\small {\bf NASA's SLS and a single-launch AEI packaging solution.}
}
   \label{fig:LV} 
\end{figure}

For the launch vehicle assessment, we assumed a 2040s launch and looked at vehicles currently in development with limited to no flight heritage. We also examined different options for a single launch of the full array. The SpaceX Starship, theoretically capable of launching 100 tons to orbit, has a fairing diameter of 7 m and excellent potential for a future packaging exercise. The NASA Space Launch System (SLS) design is currently the best solution, and there are multiple conceptual fairings possible (Figure \ref{fig:LV}). We examined the 8.4m Payload Fairing (PLF) long where the stowed AEI array fits nicely into the fairing. 

The SpaceX Falcon Heavy rocket is another option. Multiple launches using a Falcon 9 sized fairing appear to be possible with some optimization of the design of the CS. This would be launching the DS on a single rocket in one launch and then three CS, packed together, on two further launches. This could be a more cost effective method to launch the array; however, reaching the in-orbit formation with multiple flights would need to be assessed. 

\subsubsection{MDL Trades}

We performed several trade studies (Table \ref{tab:trades}). The main trade was to look at whether any benefits would be gained for this architecture in tethering the inner pair of collector satellites. This was not recommended with respect to the formation flying due to jitter and slack issues during operation. At a distance of $\sim20$ m for the inner pair, the strain under the tether load could be large due to stretching. This could be mitigated by a tension mechanism but that would require study. The coefficient of thermal expansion for the tether material could also cause distortions in the formation flying. A temperature compensation may be possible but would be power intensive, and thus not an appropriate resource cost. Finally, a full scale tether design would be difficult to test on the ground, which adds uncertainty. We note that in-space tethering may be achievable sometime in the future with In-Space Servicing Assembly and Manufacturing (ISAM). 

\begin{table}[h]
    \centering
    \caption{Trade studies performed} 
\begin{tabular}{|| l | l ||}
    \hline
{\bf Trade} & {\bf Outcome} \\
    \hline\hline
 Single or multiple launch & 7 S/C fit in a single SLS fairing \\
 \hline
 Mechanical tether between collector &  No tether. Free-flying simplifies the \\
 pairs & array reconfiguration for new targets \\
 \hline
 Chemical or electrical propulsion for  &  EP (Hall thrusters) chosen. Electrospray \\
 microthrusters on CS & required too much time to reformulate the array \\
 \hline
 Radio frequency (RF) or optical &  Optical comm chosen for security, lower contact \\
  communications for downlink & times (5 min vs. 30 min). May allow for shorter \\
  & daily maintenance periods.\\
\hline
    \end{tabular}
    \label{tab:trades}
\end{table}

Another option would be a more rigid truss system for the CS pairs. Such structures have been demonstrated at lengths of 300 m for large deployable structures for space antennas and on-orbit
structural metrology and compensation of such structures \citep{lane2011}.  We did not study how this could work but this could be the subject of a NIAC Phase II or other future study.

\subsection{Attitude Control and Formation Flying}
\label{attitudecontrol}

One positive aspect of the AEI architecture with regards to attitude control is that only the pairs of collector mirrors that form a baseline require an accurate registration with one another and so each CS pair, plus the DS, acts as its own “telescope”. To maintain the overlap of the two arms of each interferometer for the outer two CS pairs, the collecting mirrors need to maintain a pointing accuracy of $0.5''$. For the inner CS pair, since the overlapping beam width is smaller, there is a more stringent requirement of $0.1''$ accuracy. Knowledge of the pointing direction of each will need to be at least 10 times better. The 
stability requirement for the CS during an observation is milli-arcsecond (tip/tilt). 

The stringent pointing stability requires significant mitigation of jitter and microvibrations in the spacecraft. Reaction wheels create too much jitter and so we looked at different types of thrusters, which are becoming more common and have become an obvious technology need for future large astronomical observatories (see e.g., Dennehy \& Alvarez-Salazar 2018)\footnote{\href{https://ntrs.nasa.gov/citations/20180006315}{https://ntrs.nasa.gov/citations/20180006315}}. We require both large slews and fine-pointing. To perform large angle slew maneuvers we chose  traditional reaction control system (RCS) thrusters that use hydrazine. We chose microthrusters as the sole actuator for fine pointing in order to maintain focus on targets during science observations.

\subsubsection{Microthrusters and Stabilization}
\label{thrusters}

Microthruster technology has reached high TRL and is being used now commonly for space applications.
Microthrusters produce forces from a fraction of a micronewton ($\mu$N) to 10 millinewtons (mN) and they are able to counteract tiny environmental disturbances on spacecraft. Cold-gas microthrusters use a gaseous propellant while colloidal microthrusters use a process called electrospray. Electropsray works by applying a high electric potential difference to charged liquid at the end of a hollow needle. A stream of tiny, charged droplets is emitted and these generate the thrust.

Both cold-gas and colloidal microthrusters were flown on and were part of the drag-free control system (DFCS) on the NASA ST7/ESA LISA Pathfinder technology demonstration mission. Dennehey et al. (2020) found that using cold-gas or colloid micro-thrusters as the method of control actuation improves fine-pointing performance by roughly an order of magnitude compared with the Hubble Space Telescope (HST), which uses reaction wheels for attitude control. Microthrusters thus reduce the cost and technical risks of achieving a demanding pointing stability performance. Cold gas microthrusters are flight tested and have been used for attitude control and fine pointing on three recent European Space Agency (ESA) missions: Microscope (Cipolla, et al. 2016), LISA Pathfinder (LPF) (Morris, et al. 2013), and the Global Astrometric Interferometer for Astrophysics (GAIA) (Jarrige, et al. 2014). Colloid micro-thrusters (Busek Co.) are also flight-tested and flew on LPF.

\begin{wrapfigure}{L}{0.45\textwidth}
\centering
 \includegraphics[width=0.43\textwidth]{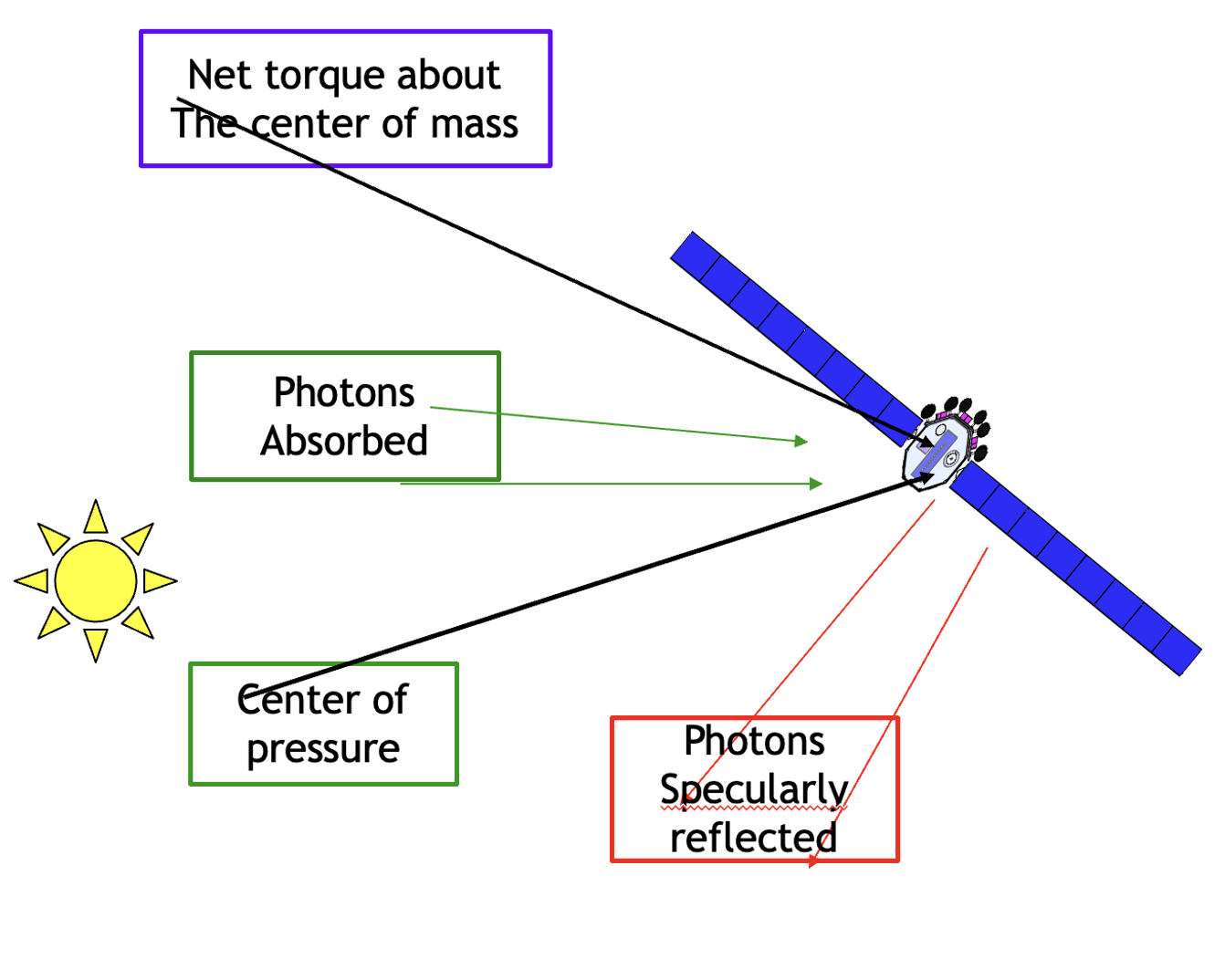}
    \caption{\small {\bf Torque from solar radiation pressure (SRP).} In the case of AEI, SRP exerts forces roughly equivalent to the required pointing stability.
}
   \label{fig:torque} 
\end{wrapfigure}

To hold the AEI spacecraft array in proper formation, the attitude control system was sized to achieve the required pointing accuracy, knowledge, and stability. The attitude control system will provide three-axis stabilization against ambient forces such as solar radiation pressure (SRP, Figure \ref{fig:torque}) and high-precision formation using star trackers (ST), an inertial measurement unit (IMU), a reaction wheel assembly (RWA), and thrusters. The thrusters and reaction wheels were sized to achieve a disturbance-free attitude control system (DFACS) and maneuverability. The RWAs for the DS were specifically sized to manage satellite disturbances during science mode. 

 Each satellite contains a sensor suite made up of Sun sensors, a star tracker system, plus a scalable space inertial reference unit (IRU) and accelerometer.  The DS carries four Honeywell reaction wheels and eight hydrazine thrusters. Each CS contains eight hydrazine thrusters and eight electric propulsion thrusters. The electric thrusters react to environmental disturbances during science observations but can also be used for slews during reconfiguration. The layout of the thruster configuration on the spacecraft ensures six degrees of freedom (dof).

 Pointing the solar arrays normal to the sun uses the Coarse Sun Sensor and IMU with the electric and hydrazine thruster actuators. 
During the science mode, fine inertial pointing is provided using the STs, science instruments, and the IMU. To unload disturbances as they come in, the DFACS uses the ST, IRU, and the electric thrusters.
The RWA/electric and hydrazine thrusters are then used for maneuvers. For fine pointing, we have the ST, IMU, science instruments (X-ray tracking with the X-ray finder scopes) and electric propulsion thrusters for disturbance free six dof attitude and extreme position control. 
  

\subsubsection{Fine Pointing Control Accuracy and Knowledge}
\label{finepointing}

Critical to the performance of AEI is attaining and maintaining pathlengths at 100 picometer knowledge ($1\times10^{-10}$ m; $1.0\times10^{-5}$ deg = 36 mas; $1.0\times10^{-7}$ mm). Pointing stability disturbances can be caused by many things, such as reaction wheel imperfections, electric thruster misalignments on the CS, environmental disturbances from SRP (see Figure \ref{fig:torque}), articulation mechanics, low-g fuel sloshing motions, and sensor noise from potential high-frequency noises in the ACS that can result in fake motion. Drift mitigation strategies for these affects include payload isolation from the bus, payload control system design for fine control, and optimizing the thruster pulsing during science observations.  

\begin{figure}
\centering
 \includegraphics[width=0.49\textwidth]{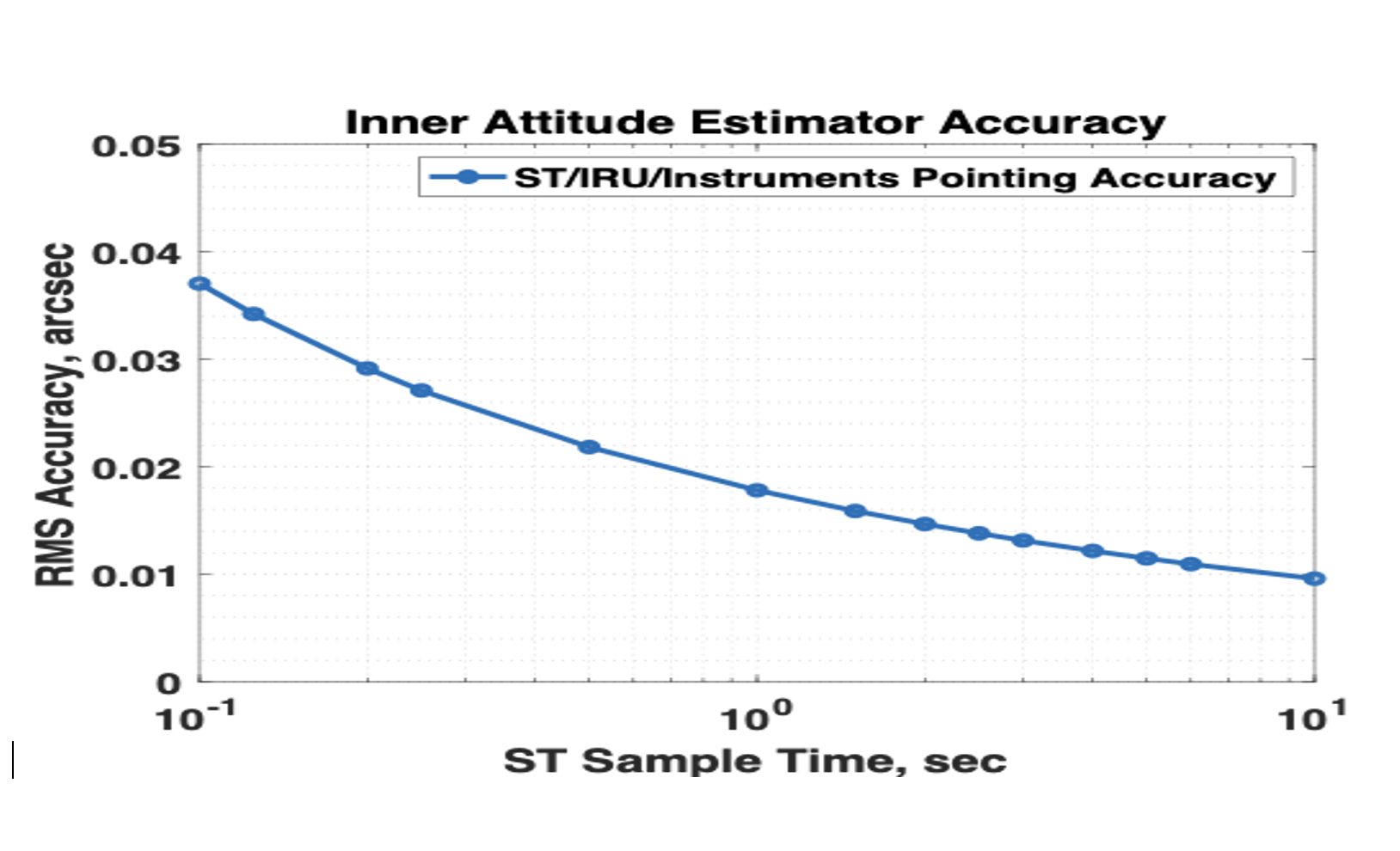}
 \includegraphics[width=0.47\textwidth]{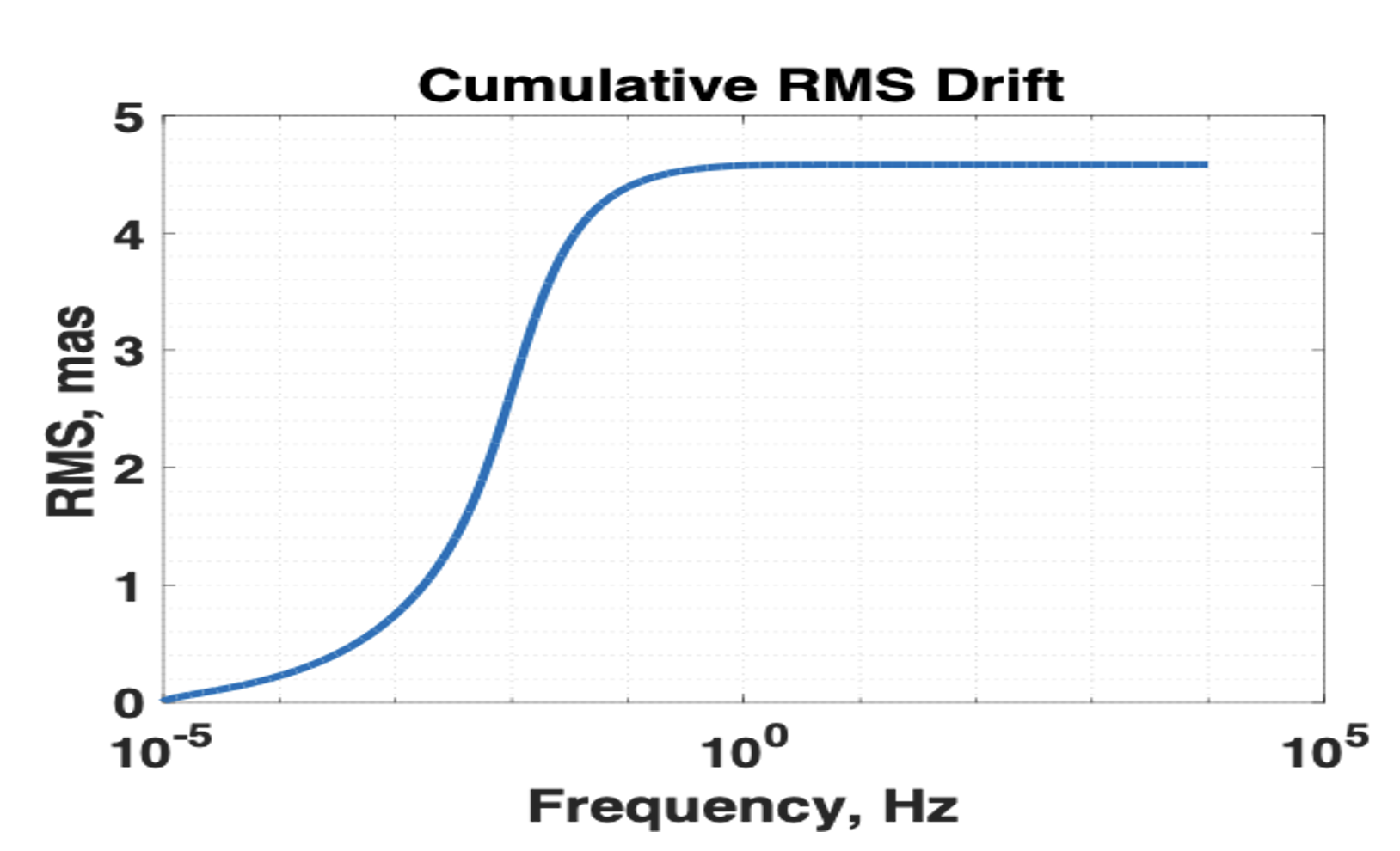}
    \caption{\small {\bf Fine pointing.} Fine pointing accuracy as a function of time and cumulative RMS drift for the inner pair of CS.
}
   \label{fig:rms} 
\end{figure}

We have determined 8.5 days on average for a science observation (see Table \ref{tab:deltav}).\ The disturbance torque from SRP is about $4.9121\times10^{-5}$ N$\cdot$m, with the ACS electric thrusters canted 60 degrees from horizontal to minimize misalignments while still producing adequate torques for mitigating disturbances. Using 13 mN thrusters for disturbance-free actuation, the set of four thrusters on a CS provides about 52 mN of thrust with a moment arm of about 1.0 m. So, on average, this means we should be pulsing every 10 minutes during a science observation to maintain the knowledge and counteract drift (Figure \ref{fig:rms}).

\subsection{Concept of Operations}
\label{operations}

\subsubsection{Mission Phases and Transfer Trajectory}
\label{transfer}

The AEI mission would consist of five phases: (1) Launch and early operations, which consists of first acquisition, attitude acquisition, and outward trajectory. The cruise time to orbit insertion is one month. Near-Earth commissioning occurs early during cruise and includes the spacecraft and instrument checkout, calibration, and validation. (2) Orbit insertion. (3) Science commissioning and calibration: fine-tuning the constellation, calibrating measurement systems. (4) Nominal operations: $2+$ years. Key to this mission is that all science targets will be planned in advance to control $\Delta$$v$ and to ensure the Sun is always located behind the target. (5) Disposal: passivation, no disposal burn.

\begin{wrapfigure}{r}{0.4\textwidth}
    \centering
    \includegraphics[width=0.38\textwidth]{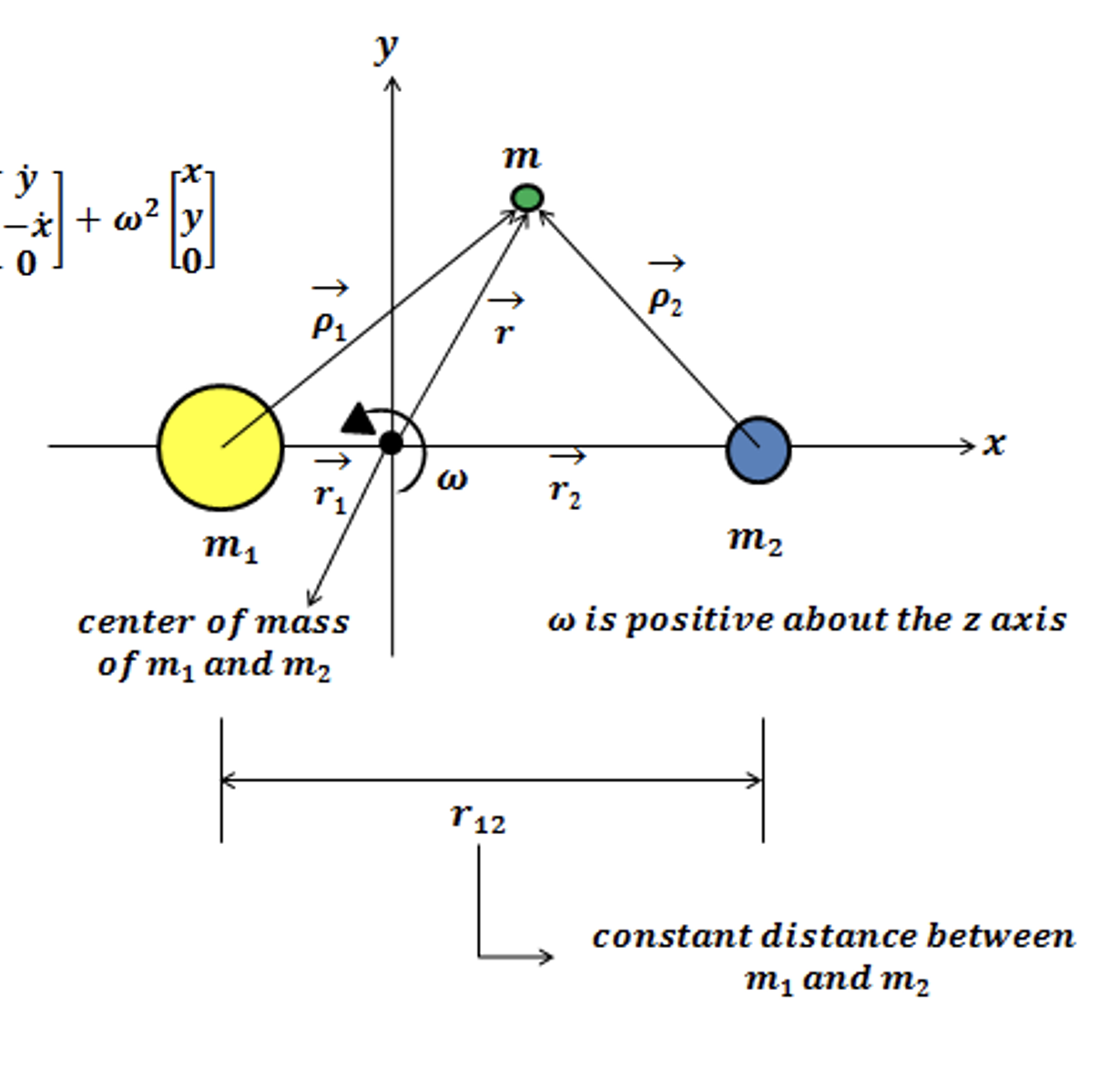}
    \caption{\small {\bf Three body problem.}}
     \label{fig:threebody} 
\end{wrapfigure}

 Planning the transfer to orbit for the interferometer starts with the required relative geometry of the seven satellites (Figure \ref{fig:constellation}). For this geometry we selected a Sun-Earth L2 halo orbit, which is at a gravitationally balanced position in space located 1.5 million km ``behind'' Earth, opposite the Sun. This is the location of other deep space observatories such as JWST and ensures that the transfer concept of operations would be in line with previously flown libration point missions. The rotating libration point frame, which we use here, is commonly used as a frame of reference to describe the dynamical motion in the circular restricted three-body problem (Figure \ref{fig:threebody}) about the Lagrange points. 

\begin{figure}
\centering
 \includegraphics[width=0.9\textwidth]{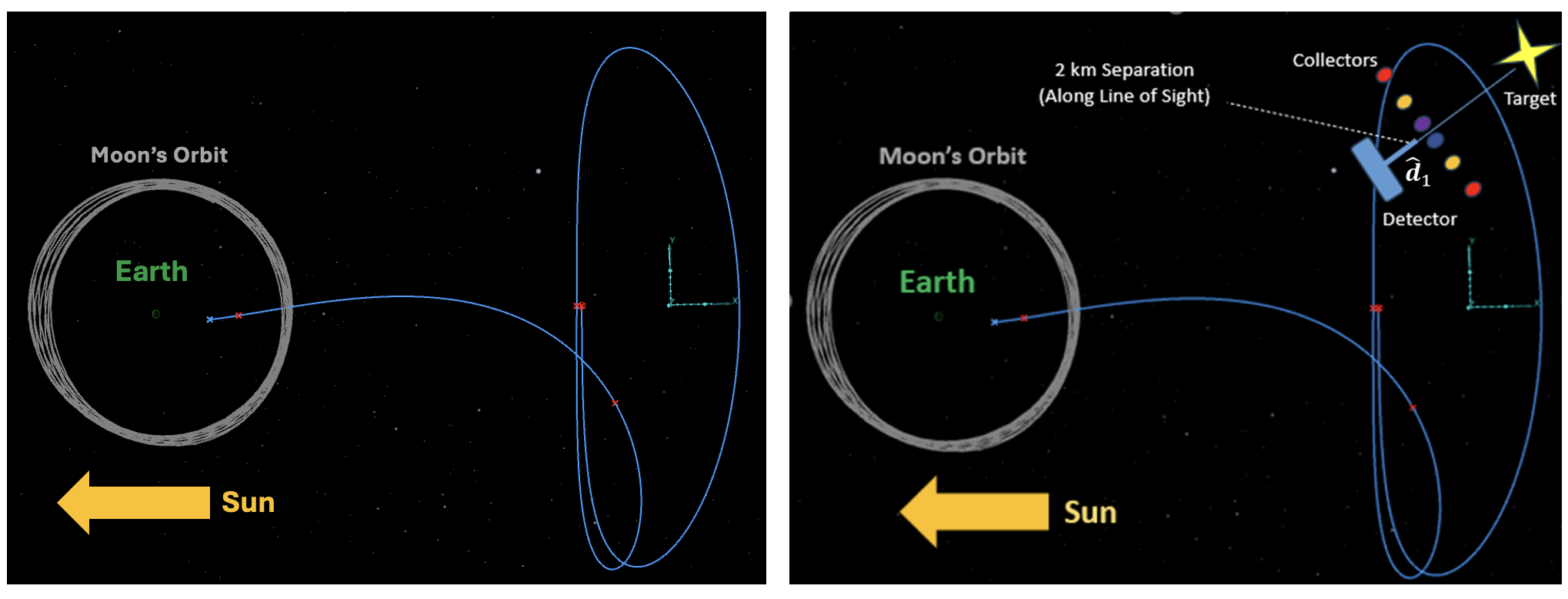}
    \caption{\small {\bf Transfer and orbit.} The orbit traces are to scale. This a Roman Space Telescope-like size Halo orbit. It takes about a year for the formation to perform a halo orbit.
}
   \label{fig:operations} 
\end{figure}

The transfer sequence is represented in Figure \ref{fig:operations}. The launch vehicle first performs the burn that directs the spacecraft towards the L2. There are two statistical Mid-Course Correction (MCC) maneuvers at the right location at the right time to fine-tune the L2 transfer trajectory. Then at the proper location, the spacecraft performs the Libration‐Point Orbit Insertion (LOI) maneuver to achieve the desired halo orbit. The halo orbit size must be optimized with respect to the pointing geometry for the target sites, the communication geometry, and eclipse events. Prior knowledge of the astronomical targets (Table \ref{tab:deltav}) plays a key role in selecting the orbit. Total transfer time is around one month from launch to orbit insertion.

\subsubsection{Targets, Formation Reconfiguration, and Observing Mode} 
\label{targets}

Once in the proper orbit, the detector slews to the required orientation with respect to the target and performs station-keeping maneuvers while the CS reconfigure to establish the desired relative geometry along the line of sight. For formation reconfiguration, the DS again follows its halo orbit and slows to the required orientation with respect to the target while the CS reconfigure to establish their relative geometry along the line of sight. Once the CS have finished observing a target, they will move back to their original halo orbit and drift naturally along the DS. Since the CS stay within a 2 km radius of the DS starting from a similar halo orbit, only very small reconfiguration maneuvers (1-2 m s$^{-1}$) are needed, excluding the $\Delta$$v$ required to match the DS’s inertial velocity for maintaining the formation.

For the AEI formation, the desired relative motion of the seven satellites is unnatural. Thus an observation will require continuous thrusting to maintain the geometry. To do that, we need to match the initial velocity of the DS throughout the observation. The total $\Delta$$v$ budget through MCCs, insertion and station keeping, including some contingency maneuvers and decommissioning, is 120 m s$^{-1}$. For our study target list (Table \ref{tab:deltav}), the total minimum and maximum $\Delta$$v$ is 399 and 638 m s$^{-1}$, respectively. This is large for a space mission lifetime and remains a tall pole with our design. A large $\Delta$$v$ requires a large reserve of propulsion. 

To compile the target list in Table \ref{tab:deltav}, ``science pillar'' targets were drawn from the three science cases developed for the MDL study. 
Targets were selected and scheduled as to minimize slewing overheads, and with consideration of the solar avoidance angle. In order to approximate targets that may be of interest to science cases that were not explored in the MDL, potential ``General Observer'' targets were identified related to AGN feedback (e.g., NGC 5728) and nuclear obscuration/AGN torus physics (e.g., UGC 11397), as well as targets with significant multi-wavelength data-sets (e.g., M87$^*$, Circinus) for which $\mu$as X-ray imaging would have legacy value.

\begin{table}[h]
    \centering
    \caption{  } 
    \begin{footnotesize}
\begin{tabular}{|| l | c | r | c | c | c ||}
    \hline
    \multicolumn{6}{|c|}{{\bf Study Target list and $\Delta$$v$ budget per CS spacecraft}} \\
    \hline\hline
    Target & Flux & Exposure & Science & Min $\Delta$v & Max $\Delta$v \\
     & cgs & (days) & & (m/s) & (m/s) \\
    \hline\hline
UGC03995A & 9.00e-13 & 14.26 & GO, AGN feedback & 30.8 & 49.3 \\
OJ 287 & 2.08e-12 & 10.43 & AGN binary, binary candidate & 22.5 & 36.1 \\
IC 2461 & 3.55e-12 & 7.55 & AGN accretion, accretion disk & 16.3 & 26.1 \\
NGC 3079 & 3.40e-13 & 12.58 & AGN binary, binary candidate & 27.2 & 43.5 \\
AD Leonis & 2.50e-11 & 3.00 & stellar flares, 19 flares in 5.8 days & 6.5 & 10.4 \\
NGC 3431 & 2.40e-12 & 11.04 & GO, AGN feedback & 23.9 & 38.2 \\
Ross 128 & 3.14e-13 & 3.00 & stellar flares, known flare star & 6.5 & 10.4 \\
NGC 4235 & 2.80e-12 & 9.17 & AGN accretion, accretion disk & 19.8  & 31.7 \\
M 87* & 1.64e-12 & 16.05 & black hole, compare with EHT & 34.7 & 55.5 \\
NGC 4939 &3.65e-12
 & 7.57 & GO, AGN & 16.4 & 26.2 \\
4U 1344-60 & 3.58e-11 & 13.25 & AGN accretion, image X-ray corona & 28.6 & 45.8 \\
Circinus Galaxy & 1.60e-11
 & 8.53 & GO, AGN & 18.4  & 29.5 \\
Proxima Centari & 8.40e-12 & 5.00 & stellar flares, major flares, $\sim$1.5/day & 10.8 & 17.3 \\
NGC 5728 & 1.33e-12
 & 6.52 & GO, AGN & 14.1 & 22.5 \\
NGC 5899 & 5.10e-12
 & 7.55 & AGN accretion, accretion disk & 16.3 & 26.1 \\
Barnard's Star & 4.80e-14 & 5.00 & stellar flares, rare large flares & 10.8 & 17.3 \\
Ross 154 & 5.68e-12 & 4.00 & stellar flares, frequent X-ray flares & 8.6 & 13.8 \\
UGC 11397 & 4.60e-12
 & 2.88 & GO, placeholder & 6.2 & 10.0 \\
2MASXJ20005575 & 6.30e-12 & 5.19 & GO, AGN feedback & 11.2  & 18.0\\
ESO234-IG063 & 7.00e-13
 & 18.93 & GO, AGN feedback & 40.9 & 65.4 \\
NGC 7172 & 2.11e-11
 & 5.04 & AGN accretion, accretion disk & 10.9 & 17.4 \\
MRK 0915 & 1.18e-11
 & 7.78 & AGN binary, binary candidate & 16.8 & 26.9 \\
\hline
Total & N/A & 184.30 & &  $\sim399$ m/s & $\sim638$ m/s \\
\hline
    \end{tabular}
    \caption{Nominal target list, exposure times, and $\Delta v$ from the MDL study. Fluxes for stellar and AGN/galaxy targets are 0.3-2 keV and 2-10 keV, respectively. Reported fluxes for stellar targets are for quiescent periods - during flares fluxes can be enhanced by two or more orders of magnitude.}
    \end{footnotesize}
    \label{tab:deltav}
\end{table}

For the stellar science case, nominal exposure times were set to 3-5 days, designed to maximize the chances of seeing at least $\sim2-3$ flares. Exposure times for the accretion and binary science cases were set based on a minimum sensitivity or surface brightness depth required for the science, assuming an effective area comparable to that of Chandra ($\sim100$~cm$^{-2}$). In short, X-ray fluxes were taken from previous studies with Chandra and XMM-Newton. As these fluxes were integrated over a much larger area than the $\sim\mu$as resolutions that AEI would be probing, we used multiple strategies to approximate what that flux might look like at these smaller spatial scales. For example, results from X-ray reverberation mapping studies were used to approximate the size and luminosity of emission coming from the innermost edge of the accretion disk. Additionally, rough theoretical simulations and results from previous studies were used to approximate what fraction of an integrated X-ray luminosity is typically concentrated into a given region (e.g., accretion disk, broad line region, etc.). From there, we determined the count rate per resolution element, and determined exposure times needed to detect a minimum of $\sim3-5$ counts in the faintest areas of interest. We acknowledge that there are significant uncertainties in these calculations (e.g., scaling factors, morphologies, etc.), and emphasize that these should only be taken as nominal exposure times used to approximate the overheads and $\Delta v$ budget. 

\subsubsection{Maintaining the Formation}
To assess the formation in the MDL study, we worked with back-of-the-envelope calculations. However, for the exact target pointings and all other constraints that come with an L2 mission, the first requirement is to determine the optimal reference halo orbit.  Initially, we assume a 3-dof model, as is standard for other space telescopes. But to truly assess AEI, we must work our way up to a 6-dof model. Everything within AEI is tightly coupled with attitude because during the array formation, the maintenance translational motions need to be continuously checked -- more challenging both computationally and in terms of algorithm development (see Section \ref{future}).

While challenging, the flight dynamics are still achievable if the navigation requirements can be met. We have not yet thoroughly explored navigation, which presents challenges due to the tight precision required. Also, a conservative value for the instantaneous acceleration that a CS must exert to nullify the effects of gravity and SRP is 40 $\mu$m~s$^{-2}$. Our conclusion is that, to optimize this mission, the entire observation schedule should be known and mostly planned in advance to minimize $\Delta$$v$. 

How do we select which target to observe first?  There is a global mission minimum in terms of $\Delta$$v$, but finding that minimum is computationally expensive for all permutations. We thus need a first-order analytical assessment to get to “good enough” and then solve that optimization problem. Depending on where the array is in the orbit, satellites are flying at different velocities. The CS must also maintain the line of sight within a 2 km separation, so wherever they are and wherever they need to be, at every point in the orbit they'll have a different $\Delta$$v$. We could save tens of m s$^{-1}$ or more by optimizing the schedule.  

\subsubsection{Concept of Operations – A Single Observation}

Observing with a multi-craft interferometer is obviously more complex than a single telescope. A day in the life consists of 23 hours of observing and one hour of housekeeping. During housekeeping, there will be SA adjustments, a communications downlink, and then target re-aquisition, which includes time for settling. Omni S band telemetry from the 6 CS is always on and the DS is receiving telemetry for “air traffic control”.

{\bf A new target command is sent:}

* Step 1 - maneuver the DS to define the target general area within $1''$ using the optical star trackers. Maneuver to lock on to the target requires tiny maneuvers in between (minutes per image to count photons).

* Step 2: Acquire with Xspot1

* Step 3: Acquire with Xspot2 

* in parallel - Array alignment - maneuver the 6 CS

* Step 4: Lock on target

\quad –	mirror alignment via fine pointing mechanism

\quad –	TBD time to stabilize thermally

* Step 5: Sit and stare (look at the field in a TBD spiral pattern to cover $u,v$ plane)

\quad –	3 to 19 days, with daily break to correct solar array

\quad –	continuous micro thruster firing and laser ranging

\quad –	collecting mirror telemetry for post-processing corrections

* Step 6: End of science collection, move to new target

We assumed AI on-board processing for steps 1 through 3. With Steps 2, 3, and 4 there is a human in the loop. Fine pointing such as this is being done with JWST. The telescope performs a gross slew with its reaction wheels and thrusters. Then once near the target, a fine guidance sensor locks onto guide stars. The fine guidance control then sends errors to a Fine Steering Mirror that makes minuscule adjustments to the telescope's optical path. These adjustments ensure that the science target remains stabilized on the detectors. We believe we can replicate this technique. 

\subsection{Communications}
\label{comms}

\begin{figure}
\centering
 \includegraphics[width=0.95\textwidth]{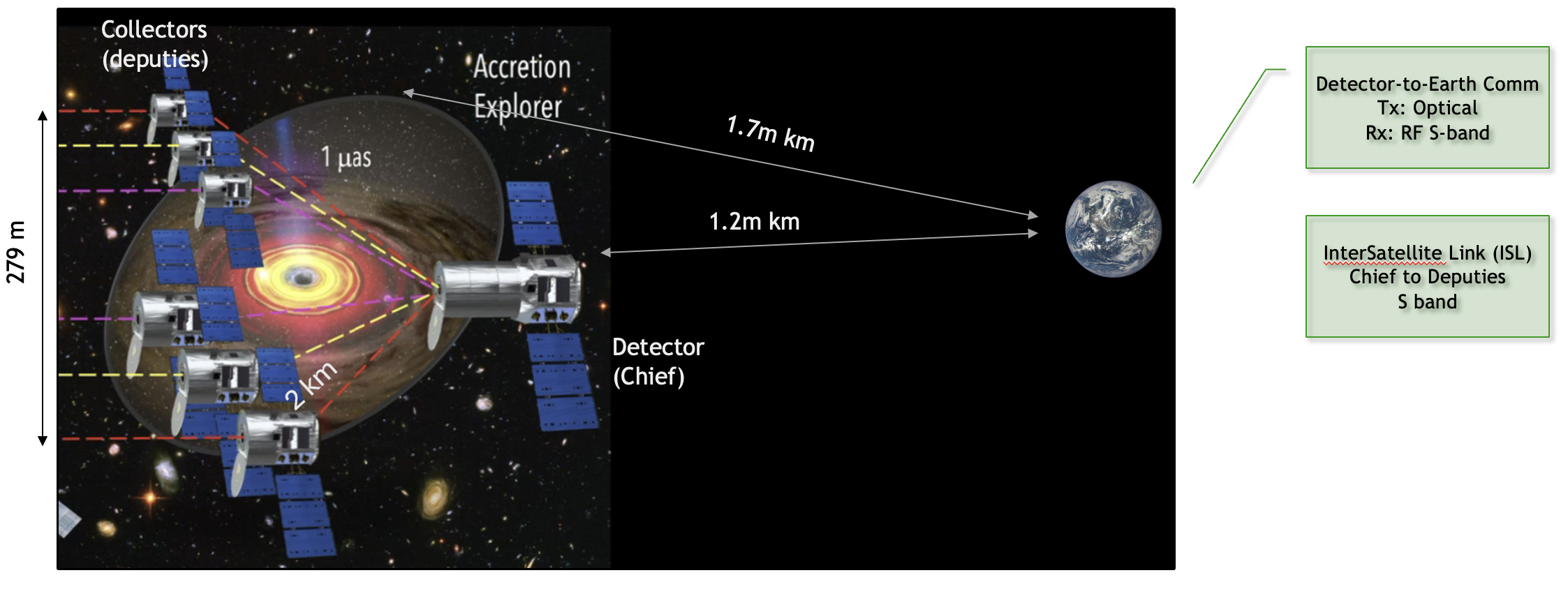}
    \caption{\small {\bf AEI Communications geometry.}
}
   \label{fig:comms} 
\end{figure}

With a fleet of seven satellites operating together, the inter-satellite and fleet to Earth communications can be complex. Deploying AEI into an L2 Halo orbit means that the minimum distance to the array is 1,223,236.4 km and the maximum distance is 1,726,605.3 km (Figure \ref{fig:comms}). For the science data download rate, we assumed the DS would transmit 5.5 Gbytes per day. We also assumed science operations would require consecutive back-and-forth Earth to DS communications during acquiring a target and there would be an inter-satellite link (DS to CS) to aid navigation in holding the relative position keeping. For operations, we examined both commanding from the Earth to the DS, and then commanding from the DS to the six CS (2 km distance). We determined that we would need direct Earth to CS commanding only for the first few days, but consider an emergency uplink channel. Finally, we assumed constant S band communication (microwave frequencies) from the DS to the CS while the microthrusters are firing -- this is a constant listening mode to make sure we understand what's happening.

\subsubsection{Science Data Transmission Trades: Radio Frequency vs.\ Optical}

\begin{wrapfigure}{r}{0.5\textwidth}
    \centering
    \includegraphics[width=0.48\textwidth]{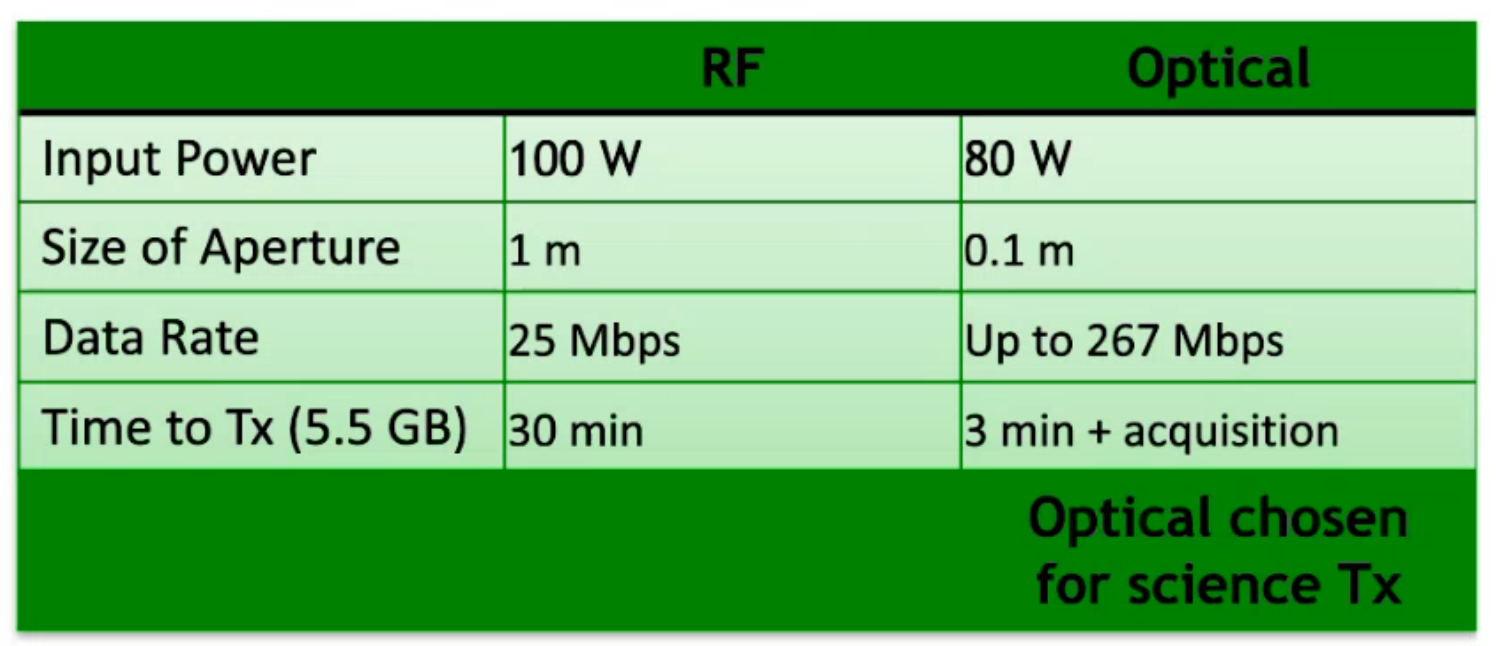}
    \caption{\small {\bf Communications comparison.}}
    \label{fig:RFtrade}
\end{wrapfigure}

We performed a trade study between radio frequency (RF) and optical communications for the transmission of science data to Earth. If we used RF, we would need a very large RF antenna. On the other hand for direct-to-Earth communications for a deep space mission, L2 is the ideal location for optical communications. A 0.1m aperture optical system 
provides an order of magnitude higher data rate than a 1m aperture RF system - up to 267 megabits s$^{-1}$ (Figure \ref{fig:RFtrade}). For the transfer of science data, the RF signal would take 30 minutes versus 3 minutes plus acquisition for optical. Also the optical communication requirement is 100 µrad pointing, much better matched to the pointing accuracy requirement for AEI. 

Optical communications is an excellent solution because it has become  more feasible for deep space with recent performance on orbit. A Modular, Agile, Scalable Optical Terminal (MAScOT) supports Low-Earth Orbit (LEO) to deep-space communication links and launched to the International Space Station in November 2023 \citep{gillmer2021}. They used a ground station at Palomar. And more recently for Artemis II, the Orion spacecraft carried an optical (laser) communications system that was the Orion Artemis II Optical Communications System (O2O). The system is capable of higher-bandwidth data transmissions and O2O used laser beams to send high-resolution video and images of the lunar surface to Earth. For its ground segment, O2O used the White Sands Complex in New Mexico and the Table Mountain Facility in California, part of NASA's Near Space Network. 

\subsubsection{Optical Communications Ground Segment}
For deep space, the ground segment is less mature now, but has time to mature by 2040. 
The DSOC Deep Space Optical Communications  experiment aboard the Psyche mission has sent data to Palomar (5m) using Pulse Position Modulation (PPM) and has successfully communicated at a distance of 140 million miles \citep{biswas2024}. Optical communications are also more secure than RF because the terminals use a narrower beam width and this provides a smaller footprint, thus dramatically reducing the area over where the communications link could potentially be intercepted. 

On 7 July 2025, ESA established its first optical communication link with a spacecraft in deep space. The link was made with NASA’s DSOC at a distance of 1.8 astronomical units, around 265 million km. A single-photon sensitive receiver for OGS is mounted to the 2.3m Aristarchos telescope, located at the Helmos Observatory \footnote{\href{https://helmos.astro.noa.gr/en/}{https://helmos.astro.noa.gr/en/}}. SSC has a presence in Dongara AU, with an Optical Ground Station (OGS) which will be 80 cm by March 2026. For uninterrupted line-of-sight coverage, two OGS are needed that are seasonally de-coupled. There is coverage of two OGS in the Americas and AU with regards to L2. This provides a 19-hour coverage, with a 3-hour overlap when two stations are visible and the best destination can be chosen based on the weather. 

\begin{figure}
    \centering
    \includegraphics[width=0.99\textwidth]{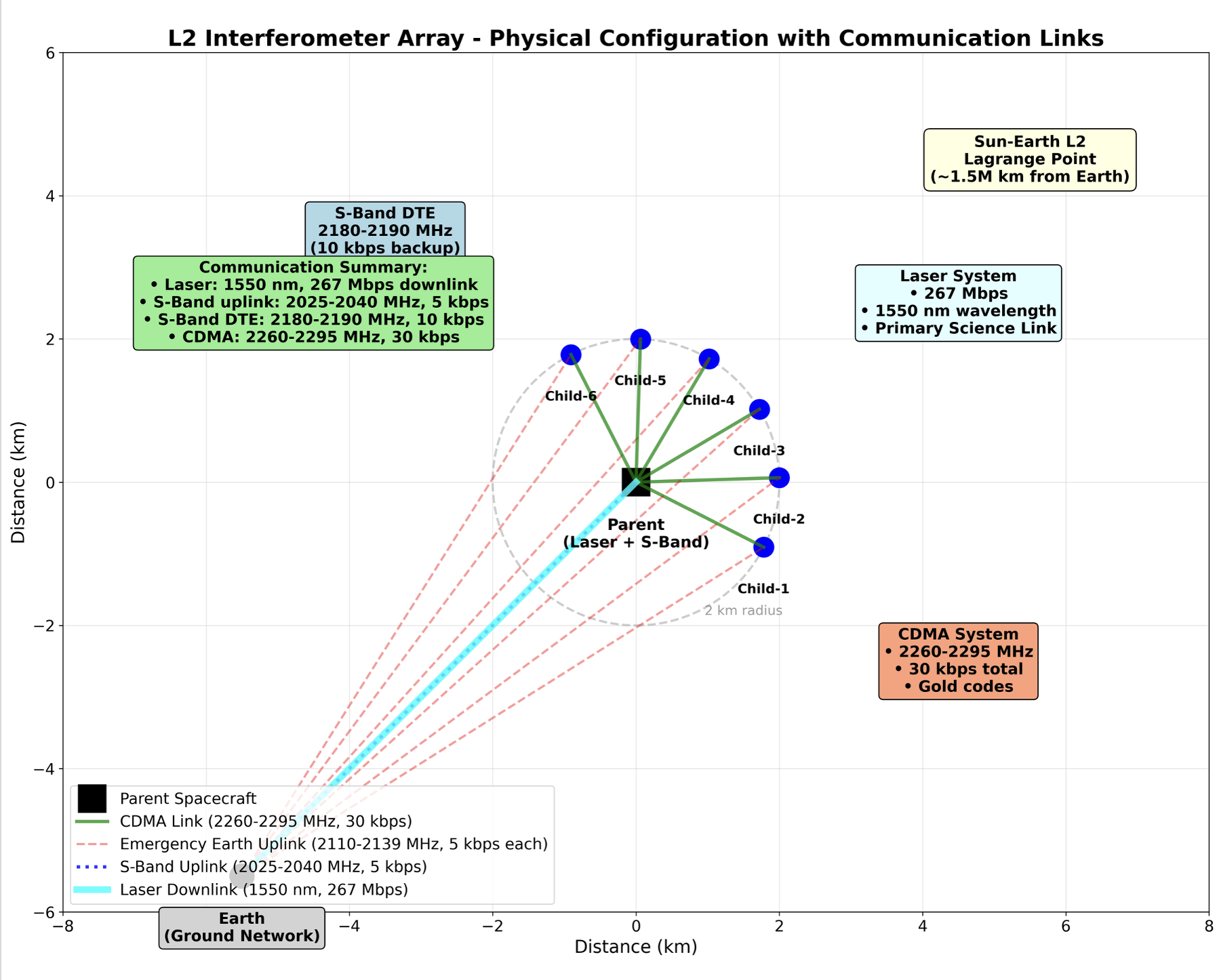}
    \caption{\small {\bf Communications spectrum for AEI.}}
    \label{fig:comms_spectrum}
\end{figure}

For AEI, we would need to start early in terms of spectrum planning because there are code-division multiple access (CDMA) links across many frequencies (Figure \ref{fig:comms_spectrum}). CDMA means that several transmitters can send information simultaneously over a single communication channel. There would be emergency links from the DS to the CS and further emergency links from the beginning of the mission for the CS to Earth and from the DS to earth. We critically will need the RF to set up the optical; pointing the RF first and get a good angle and then the optical will trace off that angle. Finally, there is a nominal link to send commands out from Earth. 

\subsection{Thermal Design}
\label{thermal}

The satellite thermal design relies on early-phase, high-level estimation techniques. The goal is to have the satellites be cold biased wherever possible. RWAs may require operational heaters to maintain their minimum operational temperatures. Also, the design includes thermoelectric coolers (TECs) for lasers, and (likely) CCD array temperature control. Pointing requirements are 
+20° to +90°, -20° to -90° over 6 months of the year and
+90° to +160°, +200° to +250° over the remaining 6 months (Figure {\ref{fig:avoidance}). The radiation environment is similar to that of the Roman Space Telescope. Laser stability is the same as the  requirements and the CCD stability and temperature control ranges are above the mK range. 

We assumed a maximum operating temperature for all components of +35 degrees C and a minimum survival temperature of -20 degrees C to -40 degrees C (depending on the component). There are liens built in for thermal distortion and sunshield design. The thermal distortion limits for the primary optics may drive tighter temperature requirements with impacts on the power budget.
Current thermal estimates do not include the CCD array temperature control systems. However, a radiator sharing, avionics bench, or other similar concept could be used to “resource share” survival power for components with comparable survival temperatures. The sunshield design for AEI requires continued concept development. The “barrel roll” of the CS for (u,v) plane coverage will complicate the design of the sunshield given the large FoV cone to be covered over the mission lifetime. The complication is that a single/dedicated “cold” side of space is not available at all Sun angles that are required for minimum science criteria.

\section{Feasibility and Concept Maturity Level}

Concept maturity levels represent a framework that can be used to measure and communicate the maturity of space mission concepts\footnote{https://hdl.handle.net/2014/44299}. 
These range ftom CML 1 to 5.

CML 1 (``Cocktail Napkin'') - The science questions have been well articulated, the type of science observations needed for addressing these questions have been proposed, and a rudimentary sketch of the mission concept and high-level objectives have been created. The essence of what makes the idea unique and meaningful have been captured.

CML 2 (Initial Feasibility) – The idea is expanded and questioned on the basis of feasibility, from a science, technical, and programmatic viewpoint. Lower-level objectives have been specified, key performance parameters quantified and basic calculations have been performed. These calculations, to first order, determine the viability of the concept.

CML 3 (Trade Space) - Exploration has been done around the science objectives and architectural trades between the spacecraft system, ground system and mission design to explore impacts on and understand the relationship between science return, cost, and risk.

For the AEI study we have reached CML2 to CML2+ in the below categories.

\begin{itemize}
 
\item {\bf Mission Development:} Key Mission Concept Parameters and performance requirements are quantified.

\item {\bf Spacecraft:} Key Flight Element Design Parameters and Performance Requirements [SC Bus performance, like slew and settle time, Orbit, Stabilization strategy, etc.] 

\item {\bf Instrument System Design:} Driving performance requirements and unique accommodation (e.g., sunshade boom) requirements identified for each instrument. Instruments have initial prioritization.

\item {\bf Technology:} Identify technologies or items that could be construed as <TRL 6. Show credibility to get to TRL 6 (detailed tech maturation plan not required) 

\item {\bf MEL:} Identify major elements (with initial mass estimate), preferably on the appropriate MEL template. Critical parameters (size and power if critical) are addressed.

\end{itemize}


\section{Challenges / Risks}
\label{challenges}

We uncovered several challenges and risks during our NIAC Phase I study. 

\begin{itemize}

\item {\bf Science:} The current low effective areas Xspot2 and even the primary interferometer are a science risk and challenge and require further study and development.

\item {\bf Navigation:} Navigation for this architecture is a challenge due to the desired high-precision relative motions. The tightly coupled orbit and attitude dynamics make formation reconfiguration and maintenance challenging, but still achievable, if the navigation requirements can be met.

\item {\bf Mass and Power Resources:} The engineering resource estimates by subsystem are very preliminary. A detailed analysis has not been performed regarding settling times after spacecraft slews and mechanism actuations. The time needed to establish positional and thermal stability may be a significant driver that affects the time between science measurements.  

\item {\bf Thermal Deformation:} Star trackers may be offset from their nominal location affecting accuracy. Reaction wheel assembly misalignments can affect the momentum and torque capacity along each of the axis. Thruster misalignments may cause pointing and positional offsets while firing (minimized with on orbit calibration).

\item {\bf CCD arrays:} The CCD array cooling may not be able to be achieved passively based on current packaging constraints.
Active cooling (i.e. TEC or Cryocooler) may be required.

\item {\bf Jitter:} The excitation of high frequency modes may induce instability or jitter. For attitude control, there are additional jitter sources such as optical communications being on during science observations and RWAs during calibration mode.


\end{itemize}

\section{Evolving the facility}
\label{evolving}

There are several ways to evolve the facility. 

\begin{itemize}

\item {\bf Add CS for wavelength coverage:}
Reaching energies down to 0.5 keV with significant area coverage requires larger grazing angles and thus larger mirror baselines (see Figure \ref{Fig:energy_bands}). But this can cover key emission line features of Oxygen and Neon which diagnose the presence of photoionized plasmas created by black holes. An extra pair of CS with a larger baseline could be added to the array.

\item {\bf Add CS for (u,v) plane coverage:} One can think of adding identical sets of CS pairs but at orthogonal angles to the current set. The (u,v) plane is meant to represent the dimensions on the sky that the interferometer covers. With pairs of CS, we obtain two dimensional fringes, but with 4 CS per grouping we could obtain more information and better image quality (Figure \ref{fig:uv}).

\begin{figure}
    \centering
    \includegraphics[width=0.5\textwidth]{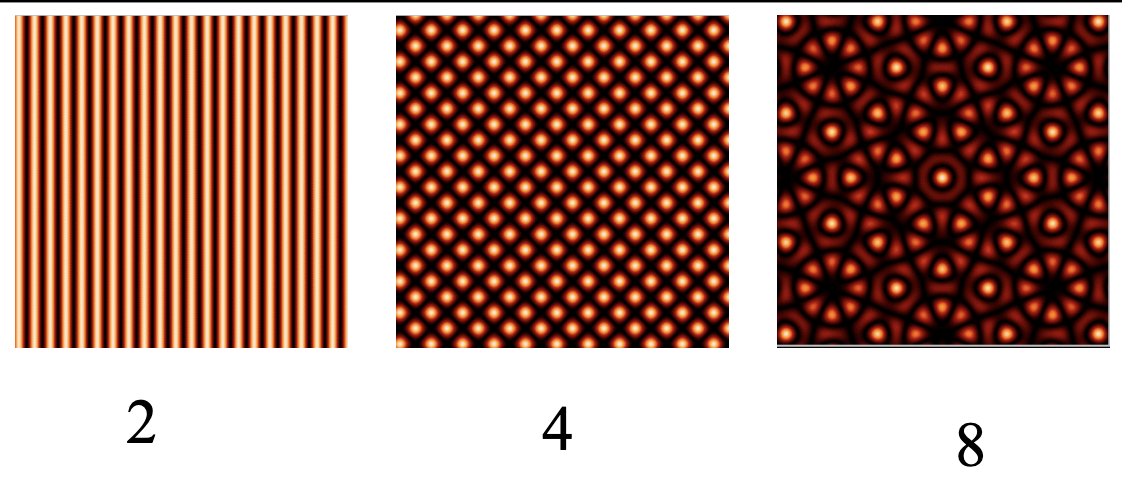}
    \caption{\small {\bf Adding mirrors at orthoginal angles.} Adding two or more collector satellites would provide better image quality. Figure from Cash NIAC Phase II report\protect\footnotemark}
    \label{fig:uv}
\end{figure}
\footnotetext{\href{https://www.niac.usra.edu/files/studies/final_report/389Cash.pdf}{https://www.niac.usra.edu/files/studies/final_report/389Cash.pdf}}

\item {\bf Non-identical CS:} For simplicity our MDL design considers identical CS. We could achieve larger collecting power at energies of 6-7 keV by making the inner CS pairs larger and holding multiple large mirror flats.  This would require further study to examine how having CS pairs be different in their design would affect attitude control and the concept of operations.

\item{\bf Modify the architecture:} Change from pure free fliers to connecting groups of satellites. We decided that tethering the inner pairs was not a solution to issues with tension and slack, but there may be a way to connect all CS with a strong material to hold them in place so that the ACS and (u,v) plane coverage of the array could be simplified. Thinking decades ahead to the 2040s, would structural connections between CS ultimately have more engineering and mission robustness than free-flyers? This could be the result of future studies.

\end{itemize}

\section{Technology Development}
\label{technology}

For the assumptions built into our study, except for the X-ray telescopes themselves, there are no low TRL items on satellite bus. Everything that is TRL 6+ except for the LISA telescopes at TRL 5+. However, the LISA capability (picometer stability) is overkill and a lower power laser system could be developed to minimize the power resource need. Potential concepts include the Laser-Ranging Interferometer (LRI) on GRACE-FO. The system  tracks a 300 km distance to nm precision (25kg, 35W).  We also note the development of the Relative Position Sensing System for Precision Formation Flying Spacecraft (PI: Anne-Marie Novo-Gradac, TRL4). This is an integrated package of optical sensors and guidance, and navigation and control software. 

Another technology development area is driven by the propulsion situation. For our needs, the electrospray microthrusters don't produce quite enough power and require too much time to reformulate the constellation. Our feasibility was shown with TRL 6+ microthrusters, but development of a ``midsize'' solution between the Hall thrusters (mN) and the Electrospray ($\mu$N) would simplify the attitude control system.

We also need to further refine the fine pointing mechinism for this array. Even with the LISA telescope adaptation, we need to be able to achieve and maintain the line-of-sight alignment for $\mu$as spacing and maintaining fringes on the detector array.

\section{Future Work}
\label{future}

While our initial feasibility assessment was successful, our study has highlighted several avenues for future work regarding the AEI concept.

\begin{itemize}

\item {\bf Target Acquisition and Centering:} We need to develop a proof-of-concept for the target centering methodology and handoff between the star tracker, Xspot1, Xspot2, and X3. This includes a proof-of-concept for laser ranging and the command methodology for formation flying using Autonomous Navigation, Guidance, and Control (AutoNGC) and the spotting telescope inputs. An end-to-end radiometric analysis is needed to determine the resultant science signal on the detectors. And finally, we need to determine the necessary settling times to re-acquire a target and re-establish the desired positional and thermal stability for an observation after each actuation or communications downlink.

\item {\bf Spotting Telescope development} This is a specific need to develop the X-ray spotting telescopes in our design. These are still very low TRL and nothing currently exists that could satify the 0.1 arcsec and milli-arcsec requirements.

\item {\bf Orbit and Attitude Control; Formation Maintenance and Reconfiguration:} New tools are needed to optimize planning and for accurate mission design assessment. To determine the optimal reference halo orbit, the size and period of the DS orbit must be determined subject to several mission constraints, including the target site geometry, the communication geometry, and eclipse events. A tool capable of applying several feedback control laws is needed that maintains the formation by matching the inertial velocity of the detector for the desired period (3-dof). Ultimately, this tool needs to be extended to 6-dof since we require active control for both translational and rotational motion during the observation mode. 

\item {\bf Propulsion:} The telescopes, deployables and radiator surfaces will need to be avoided when accounting for thruster plumage. Active thermal management may be required for electrical propulsion due to the power dissipated. This is a significant issue for the ideal placement of the LISA telesopes on the DS.

\item {\bf Structure analysis:} We require a high frequency modes analysis for the spacecraft structure to assess its impact on the attitude control system. It is important to investigate isolators for jitter suppression. Also, we need a trade study between isolating the  payload \& spacecraft versus individual isolators to isolate components.
 
\item {\bf Sunshades:} Evaluate deployable and fixed sunshade options for the top of the DS and the top and bottom of the CS. It would be good to avoid complex soft deployables if at all possible. The Roman Space Telescope uses a soft deployable because of limited space in the launch vehicle fairing. AEI may not have this issue since the spacecraft are expected to be shorter than RST. However, further work is required to develop the thermal needs for the sunshades.
 
\item {\bf Electric propulsion:} Assess the impact of Hall effect thrusters on the CS Design.  The placement of the electronics and the required power increase may require a dedicated active thermal system. For the CS disturbance free attitude we need to optimize the number of millithruster pulses during science observations. A trade study between hydrazine and electric thrusters on CS during target reconfiguration is needed.

\item {\bf Thermal:}  Develop the next level of thermal requirements, including the overall temperatures, stability, gradient, and distortion. Fully assess the thermal environments on the spacecraft over the full pointing FoV to identify and characterize the radiator accommodation options. Develop the preliminary CCD array concept for X3, including a passive versus active thermal control assessment.

\item {\bf Electrical Design and Solar Array Requirements:} A more detailed look at SA sun angles and off-pointing scenarios is required. This is especially important given the need to rotate the array for (u,v) plane coverage. SA size and margin can be refined based on better load estimates. 

\item {\bf (u,v) plane coverage:} At the mission level, we need to understand (u,v) plane sampling requirements during observations and produce tools to model the image reconstruction.

\end{itemize}

\section{Publications, presentations, collaborations}

{\bf Publications:}

Weaver K.A., et al. (2026) The need for ultra high resolution X-ray imaging. Front. Astron. Space Sci. 13:1790876. doi: 10.3389/fspas.2026.1790876 \citep{weaver2026}

Rohrbach, S., et al., 2026, SPIE, in prep (paper on the beamsplitter combiner optic)

Cann et al. in prep (a full STM for AEI)
\newline
{\bf Presentations:}
K. Weaver's High-ReX SAG Interferometry talk on June 5.
\newline
https://science.nasa.gov/astrophysics/programs/physics-of-the-cosmos/community/astra-space-interferometry-webinar-5-june-2026/
\newline
{\bf Press:}
https://www.universetoday.com/articles/watching-the-power-of-supermassive-black-holes-with-x-ray-interferometers.
\newline
{\bf Collaborations:} Our NIAC Phase I team has other collaborators outside of this Phase I study, such as the science collaborators for our white paper \citep{weaver2026}. We are also forming collaborations with external partners to examine technology development.

The Physics of the Cosmos X-ray Science Interest Group (SIG) has formed a Science Analysis Group (SAG) to examine high resolution X-ray imaging (Hi-ReX). K.A. Weaver co-leads the HiReX SAG, which provides a venue for scientific community members to develop science cases for future high-resolution X-ray imaging missions. Other AEI collaborators involved in this effort include Jenna Cann, Jeffrey McKaig, and Laura Vega.

Since completing the MDL study, we have been engaging potential collaborators for technology development relevant for AEI under the Astrophysics Strategic Technology and Reserach Accelerator (ASTRA) Initiative. This includes Xspot2 and a truss option for the CS.

\section{Conclusions}

Our Phase I NIAC study has shown that the AEI concept is feasible in operation and we have progressed the mission concept to a feasible architecture for further study to expand the possibilities around today's future X-ray missions. The type of satellite station keeping for high precision formation flying for an X-ray interferometer is now possible by leveraging off of LISA pathfinder technology and other recent developments in laser ranging. Using mirror flats plus an X-ray beam splitter for the prime $\mu$as optic is feasible. The required science modes are easily accommodated and the pointing accuracy is satisfied. The electric mN thrusters are capable of achieving a disturbance free attitude and the reaction wheel assemblies are capable of slewing the observatory, storing enough momentum for momentum management

Next steps include a proof-of-concept for target aquisition and to study how to best operate the array to achieve a minimum $\Delta$v to maximize science return. Dedicated X-ray optics development is absolutely required as a key enabling technology. However, we believe that with this development beginning now, the AEI concept could have test flights for components in 5 years, test flights for the formation on a smaller scale in 10 years and a full-scale large mission in 20-25 years.

\bibliographystyle{aasjournalv7}
\bibliography{test}

\newpage
\section{Appendix A: Additional Notional Mission Descriptions}
\label{AppendixA}

\subsection{Dual and Binary AGN, Gravitational Wave Events}
Dual and binary AGN serve as the precursors to SMBH mergers detectable with pulsar timing arrays (PTAs) or the Laser Interferometer Space Array (LISA), and are therefore can provide key insight into the merger-driven growth and evolution of SMBHs. Moreover, recent discoveries with XMM-Newton and NICER \citep[see][for a review]{kara2025} have identified semi-periodic extreme transient events (quasi-periodic eruptions; QPEs) that may be indicative of extreme mass ratio inspirals (EMRIs), where an orbiting low mass object interacts with the central supermassive black hole.
With $\mu$as resolution X-ray imaging capabilities, we can investigate questions such as: \textit{How do accretion and obscuration properties of dual and binary AGN change as a function of pair separation? How do these properties affect the "final parsec problem" to ultimately result in a BH merger? Are QPEs truly caused by EMRIs, or are they instead caused by accretion disk instability or other events?}

\begin{itemize}
\item[] \textbf{Angular resolution:} 1 – 10 $\mu$as for local mpc binaries (z < 0.05)

\item[] \textbf{Bandpass:} Fe K ($\sim6-7$ keV), broadband ``hard'' X-ray band ($> 2$ keV), Soft X-rays ($< 2$ keV); with these bandpasses, we would be able to compute hardness ratios, and determine a basic spectral shape and rough obscuration estimates

\item[] \textbf{Sensitivity:} At least $10^{-14}$ erg~cm$^{-2}$~s$^{-1}$ (lowest estimate: $1\times10^{-16}$ erg~cm$^{-2}$~s$^{-1}$); ideally this is per resolution element, but can be multiple resolution elements binned together. Significant uncertainties in this due to limited understanding of the size of the emitting region that we will be probing. This science gap could be resolved with a mas precursor.

\item[] \textbf{FOV:} $\sim100$s of $\mu$as (or a mosaicking capability)

\item[] \textbf{Spectral Resolution:} CCD quality is ideal, or rough multi-band imaging is sufficient.

\item[] \textbf{Observation exposure times:} Approximately a week or more

\item[] \textbf{Other Requirements:} Effective area comparable to Chandra would enable the detection of moderate luminosity QPEs (peak $0.05-1$ ct/s) with significant counts in $<1$ hr exposures, which is needed to probe both before, during, and after the event.

\end{itemize}

\subsection{Stellar Science and Stellar Flares}

 A spatially resolved coronal image provides valuable insight into the dynamic behavior and spatial distribution of energy release during intense flare events. Currently, the Sun is the only star for which we possess a spatially resolved coronal image. Achieving higher spatial resolution in X-ray observations is essential for extending these insights to stars across a range of masses and evolutionary stages, and for understanding how their coronal properties vary. In particular, we seek to answer science questions such as: \textit{How does the spatial distribution and morphology of stellar flare and magnetic activity affect atmospheric escape and planetary habitability?}

\begin{itemize}
\item[] \textbf{Angular resolution:} $\sim$µas to spatially resolve stellar flares for a significant population. Mas-accessible targets also available.

\item[] \textbf{Bandpass:} Soft X-rays

\item[] \textbf{Sensitivity:} To study both quiescent and flaring activity for stars, the instrument should achieve a sensitivity of $10^{-12}$ to $10^{-13}$ erg s$^{-1}$ cm$^{-2}$ with time averaging on the order of seconds to minutes.

\item[] \textbf{Observation exposure time:} 1 to 3 days 

\item[] \textbf{Other Requirements:} Timing resolutions on the order of seconds/minutes

\end{itemize}

\begin{table*}[h]
\caption{Nearby stellar targets illustrating angular diameter ($\theta_{\rm diam}$) and X-ray brightness, ordered by decreasing angular diameter (full table in \citep{weaver2026}.}
\label{tab:stellar_candidates}
\centering
\resizebox{\textwidth}{!}{%
\begin{tabular}{lcccccc}
\hline
Star & SpT & Distance (pc) & $\theta_{\rm diam}$ (mas) & $F_{X,q}$ (erg cm$^{-2}$ s$^{-1}$) & log $L_X$ & Planet Status \\
\hline
Alpha Centauri A+B (combined) & G2V+K1V & 1.35 & 8.56, 5.80 & $9.9\times10^{-12}$ & 27.33 & No confirmed planet \\
Proxima Centauri & M5.5Ve & 1.30 & 1.10 & $8.4\times10^{-12}$ & 27.23 & Known planet(s) \\
AD Leo & M3.5V & 4.97 & 0.79 & $2.5\times10^{-11}$ & 28.87 & No confirmed planet \\
AU Mic & M1Ve & 9.72 & 0.67 & $1.8\times10^{-11}$ & 29.31 & Known planet(s) \\
BD-17 588A & M3V & 6.87 & 0.37 & $1.98\times10^{-12}$ & 28.05 & Planet host/candidate (TOI) \\
AP Col & M4.5Ve & 8.66 & 0.31 & $4.8\times10^{-12}$ & 28.63 & No confirmed planet \\
\hline
\end{tabular}}
\footnotesize
Angular diameters are computed from adopted stellar radii and distances. $F_{X,q}$ values are taken directly from X-ray catalogs when available; otherwise, when only ROSAT-like count rates were available (and hardness ratios were not), fluxes were estimated assuming a fixed conversion factor $\mathrm{ECF}=1.1\times10^{-11}$ erg cm$^{-2}$ ct$^{-1}$ in the ROSAT PSPC 0.1--2.4 keV band, appropriate for typical nearby stellar coronal spectra and negligible interstellar absorption. ROSAT matches are restricted to good positional associations ($\leq 60''$).
\normalsize
\end{table*}

Note: Given the constraint we outline in Section \ref{zones} of +/-20 degrees off pointing (Figure \ref{fig:avoidance}), we will need to think carefully about stellar observations confined only to the galactic plane and near solar-neighborhood (owing to sensitivity) may be affected. This will be a concern during target cataloging phase.

\newpage

\section{Appendix B: Pre-MDL Conceptual Drawing and Baseline Calculation}
\label{AppendixB}

The conceptual drawing is shown in Figure \ref{fig:mirrorflats}, where at the very left we show a set of mirror flats that are housed within a 2.4 m diameter aperture to produce mas angular resolution. The mirrors reflect X-rays at different energies from $\sim0.7$ keV to $\sim8$ keV denoted by the solid angled lines in the adjoining panel where, from the outside in, the lowest energy X-rays reflect from the outer mirrors and the highest energy X-rays reflect from the inner mirrors. Expanding this aperture to larger sizes shown on the right produces a system that can provide 1000 times better resolution.

\begin{figure}[h]
\centering
 \includegraphics[width=0.9\textwidth]{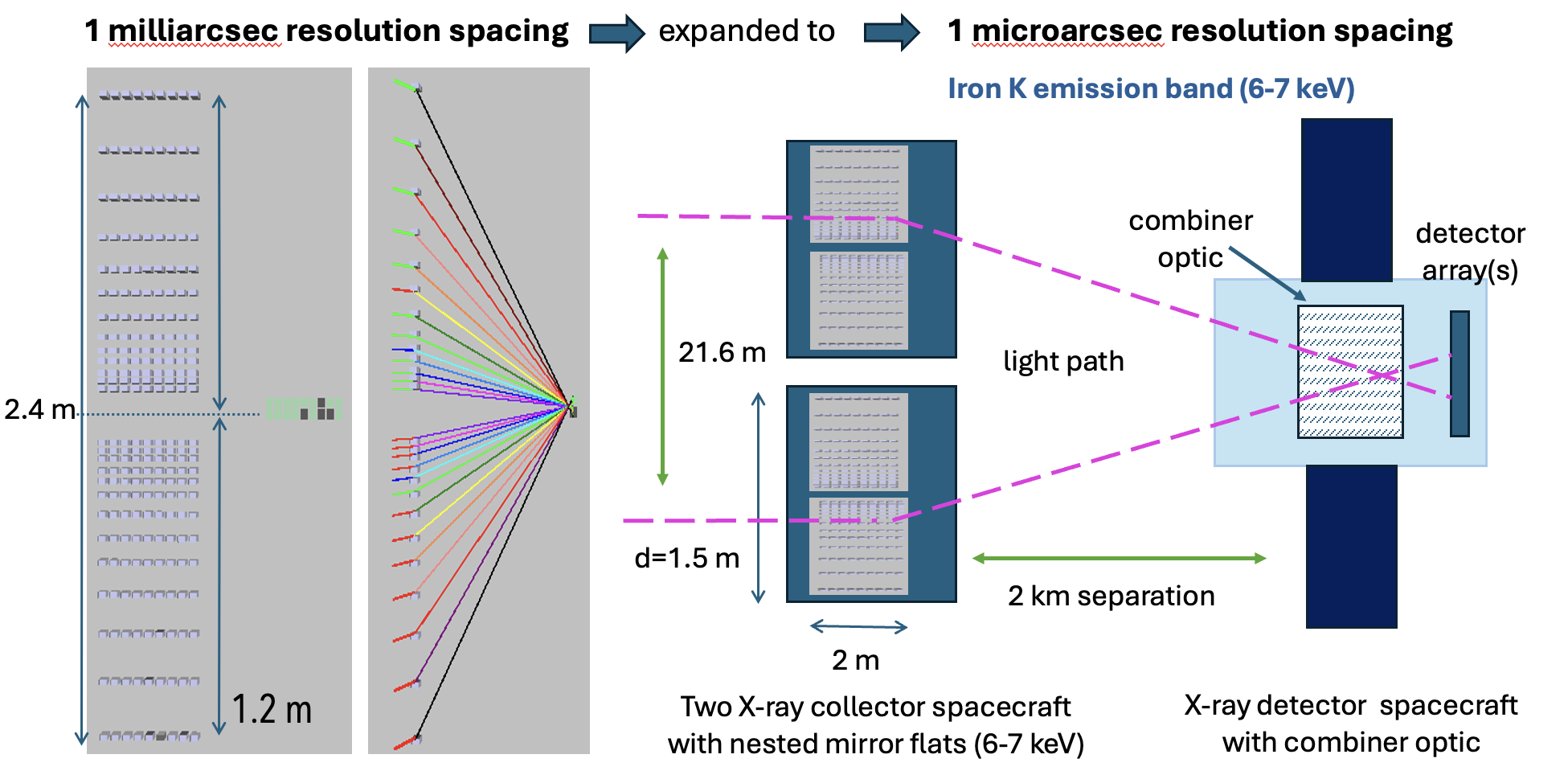}
    \caption{\small {\bf Original conceptual drawing of the dimensions for the interferometer's optical path.
    } Left: Sets of mirror flats spaced apart to fill a 2.4 m aperture providing mas resolution. Outer mirrors collect $\sim0.7$ keV photons and inner mirrors collect up to $\sim8$ keV. The second image shows the pairs of ray paths with different wavelengths as colors from longer (black) to shorter (purple). Right: Our architecture expands these mirror baselines with individual spacecraft to achieve $\mu$as resolution. A 21.6 m spacing focuses X-rays of $\sim6-7$ keV.  
}
   \label{fig:mirrorflats} 
\end{figure}

We calculate the baselines for these channels from the required resolution of the telescope at $\Theta_r$ = $\lambda$/2D. As an example, setting the lower energy bound of the inner channel at 5.7 keV (0.216 nm), then for $\mu$as resolution, the collecting mirror separation is D = $\lambda$ / 2x1 $\mu$as = 21.6 m for the inner mirror pair. The distance from the collecting to the combining optic is then 2 km $\pm10$ m.

\end{document}